\documentclass[journal]{vgtc}                
\ifpdf
  \pdfoutput=1\relax                   
  \usepackage{graphicx}                
  \DeclareGraphicsExtensions{.pdf,.png,.jpg,.jpeg} 
\else
  \usepackage{graphicx}                
  \DeclareGraphicsExtensions{.eps}     
\fi%

\graphicspath{{figures/}{pictures/}{images/}{./}} 

\usepackage{microtype}                 
\PassOptionsToPackage{warn}{textcomp}  
\usepackage{textcomp}                  
\usepackage{mathptmx}                  
\usepackage{times}                     
\usepackage{cite}                      
\usepackage{tabu}                      
\usepackage{booktabs}                  

\usepackage{color}

\usepackage{makecell}
\usepackage{diagbox}
\usepackage{setspace}
\usepackage{float}
\usepackage{multirow}
\usepackage{rotating}
\usepackage{graphicx}
\usepackage{caption}
\usepackage{url}
\usepackage{soul}
\usepackage{wrapfig}

\onlineid{1532}

\vgtccategory{Research}
\vgtcpapertype{Empirical Study}

\title{Revisiting Dimensionality Reduction Techniques for
Visual Cluster Analysis: An Empirical Study}

\author{Jiazhi Xia, Yuchen Zhang, Jie Song, Yang Chen, Yunhai Wang and Shixia Liu}
\authorfooter{
\item
 Jiazhi Xia, Yuchen Zhang and Jie Song are with School of Computer Science and Engineering, Central South University, China. E-mail: \{xiajiazhi,yuchenzhang,jiesong\}@csu.edu.cn;
\item
 Yang Chen is with I4 data. E-mail: chen1984yang@gmail.com.
\item
 Yunhai Wang is with School of Computer Science and Technology, Shandong University, China. E-mail: cloudseawang@gmail.com.
\item
 Shixia Liu is with School of Software, Tsinghua University, China. E-mail: shixia@tsinghua.edu.cn.
 \item
 Jiazhi Xia and Yang Chen are corresponding authors.
}

\shortauthortitle{Biv \MakeLowercase{\textit{et al.}}: Global Illumination for Fun and Profit}

\abstract{Dimensionality Reduction (DR) techniques can generate 2D projections and enable visual exploration of cluster structures of high-dimensional datasets. However, different DR techniques would yield various patterns, which significantly affect the performance of visual cluster analysis tasks. We present the results of a user study that investigates the influence of different DR techniques on visual cluster analysis. Our study focuses on the most concerned property types, namely the linearity and locality, and evaluates twelve representative DR techniques that cover the concerned properties. Four controlled experiments were conducted to evaluate how the DR techniques facilitate the tasks of 1) cluster identification, 2) 
membership identification, 3) distance comparison, and 4) density comparison, respectively. We also evaluated users' subjective preference of the DR techniques regarding the quality of projected clusters. The results show that: 1) Non-linear and Local techniques are preferred in cluster identification and membership identification; 2) Linear techniques perform better than non-linear techniques in density comparison; 3) UMAP (Uniform Manifold Approximation and Projection) and t-SNE (t-Distributed Stochastic Neighbor Embedding) perform the best in cluster identification and membership identification; 4) NMF (Nonnegative Matrix Factorization) has competitive performance in distance comparison; 5) t-SNLE (t-Distributed Stochastic Neighbor Linear Embedding) has competitive performance in density comparison.

} 

\keywords{Dimensionality reduction, visual cluster analysis, perception-based evaluation}

\CCScatlist{ 
 \CCScat{K.6.1}{Management of Computing and Information Systems}%
{Project and People Management}{Life Cycle};
 \CCScat{K.7.m}{The Computing Profession}{Miscellaneous}{Ethics}
}

\teaser{
  \centering
  \includegraphics[width=\linewidth]{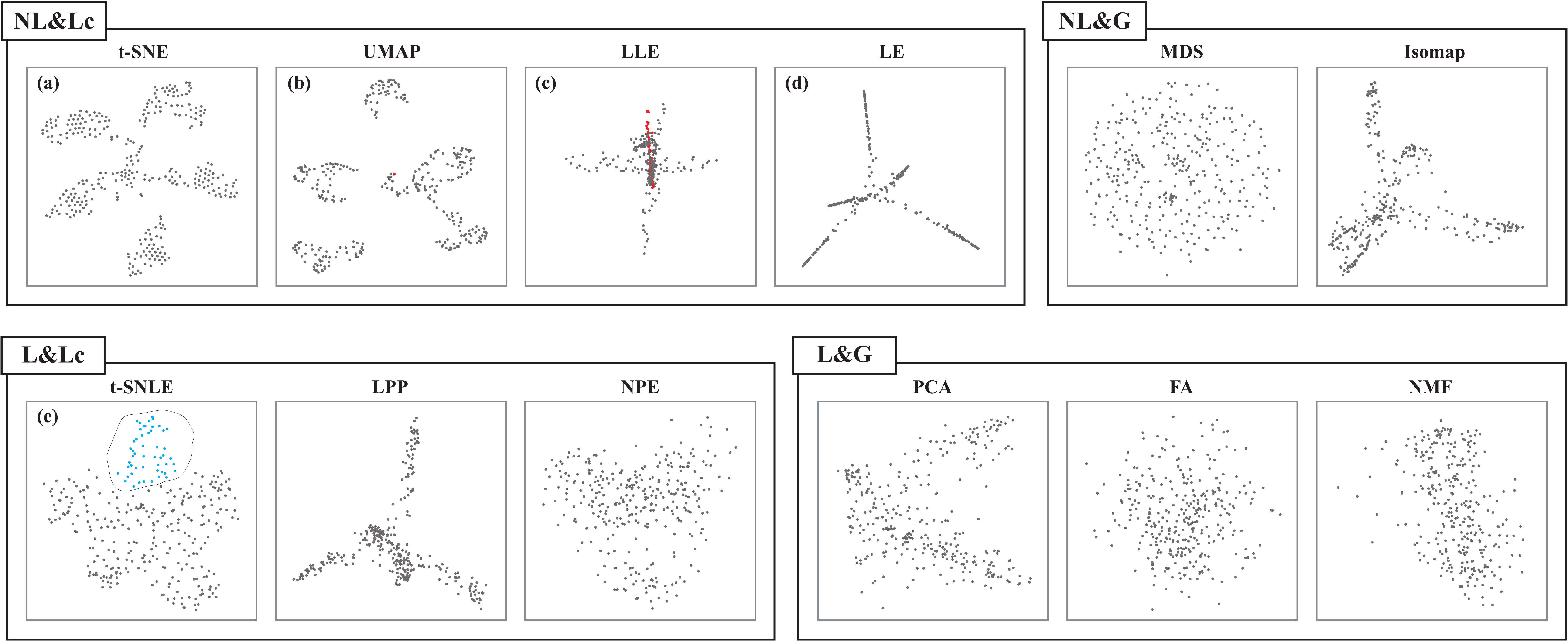}
  \caption{Different dimensionality reduction results of the ExtYaleB dataset, which contains 320 images of five people. The twelve dimensionality reduction techniques fall into four categories: Non-linear \& Local (NL\&Lc), Non-linear \& Global (NL\&G), Linear \& Local (L\&Lc), and Linear \& Global (L\&G). Instances of task stimuli: (a) Cluster identification (\textbf{T1}), (b) Membership identification (\textbf{T2}), (c) Distance comparison (\textbf{T3}), and (d) Density comparison (\textbf{T4}). (e) lasso tool used in \textbf{T1--T4}.}
  \label{fig:teaser}
}

\vgtcinsertpkg

\begin{document}




\firstsection{Introduction}

\maketitle

Visual cluster analysis usually employs dimensionality reduction (DR) techniques to project high-dimensional data into 2D scatterplots, in which analysts can visually identify cluster patterns~\cite{wenskovitch2017towards, yuan2020evaluation}. The cluster analysis is an inherent human-in-the-loop task that lacks a universal ground truth~\cite{aupetit2019toward}. Therefore, visual cluster analysis, which assumes clusters in the projection are faithful representation matching with actual clusters, is widely used in evaluating automatic clustering algorithms~\cite{aupetit2019toward} and clustering applications~\cite{becht2019dimensionality,han2019visual,yue2019sportfolio,yuan2021survey,liu2017towards}.

However, different DR techniques perform differently in a set of aspects in visual cluster analysis, e.g., cluster separation, membership preservation, distance preservation, and density preservation. As shown in Fig.~\ref{fig:teaser}, the twelve DR techniques generated various visual patterns out of the same dataset. Although each DR technique provides different insights into clusters, most of them are rarely used in real applications. For example, we have surveyed papers in the mainstream visualization journals and conferences from 2010 to 2020 with a focus on applying DR techniques in visual cluster analysis. Among the 51 resulted papers, we found that only PCA (Principal Component Analysis)~\cite{wold1987principal}, MDS (Multidimensional Scaling)~\cite{kruskal1978multidimensional}, and t-SNE (t-Distributed Stochastic Neighbor Embedding)~\cite{maaten2008visualizing} are used for more than $2$ times. A revisit to DR techniques is needed to guide the use of them for visual cluster analysis. 

Moreover, a high-level view of the performances of DR techniques on visual cluster analysis is a common concern of researchers. When choosing DR techniques from a large set, analysts usually consider the property that the DR technique preserves. Among various properties, linearity and locality have gained the most attention~\cite{nonato2018multidimensional, espadoto2019towards}. For example, linear DR techniques are believed to outperform in preserving the density of clusters~\cite{chatzimparmpas2020t,narayan2021assessing}. On the other hand, non-linear DR techniques, such as t-SNE and UMAP (Uniform Manifold Approximation and Projection)~\cite{mcinnes2018umap}, are widely used because of their performance in separating clusters. The locality property, i.e., if local or global pairwise relationships are preserved, also yields significant differences in the generated scatterplots~\cite{espadoto2019towards, de2003global}. 

Nevertheless, many qualitative comparisons have been conducted towards understanding how different DR techniques perform in visual cluster analysis tasks~\cite{etemadpour2014perception,ventocilla2020comparative, xu2017evaluating,zhao2018evaluating}. However, we argue that two significant gaps exist in current research. First, while the aforementioned high-level properties are of concern to both DR researchers and analysts, there lacks experimental studies that test and verify them against user's visual perception in visual cluster analysis tasks. Second, none of the existing experimental studies has evaluated a set of DR techniques that can be representative of all the properties or cover the most popular DR techniques, calling for a comprehensive and systematic comparison of these methods.

To fill these gaps, we conduct controlled experiments to evaluate DR techniques in visual cluster analysis. Given the same high-dimensional datasets and certain cluster analysis tasks, our study compares the effectiveness of graphical outcomes generated by different DR techniques and investigates how their essential properties affect user's visual perception and the task performance. The study starts with a comprehensive literature review of existing DR techniques, upon which twelve widely-used techniques are selected for evaluation (see Table \ref{tab:dr}). They are grouped based on two properties, namely locality and linearity, both of which are among the most concerned topics in the existing DR research. This grouping allows a high-level view of the DR techniques regarding their core features and thus the study results can be easily understood by data analysts. To ensure the study results are intuitive and generalized to various domains, we carefully define four analytical tasks that are essential for general visual cluster analysis. Each task is evaluated through a controlled experiment. The four corresponding controlled experiments are assessed based on ground truth established in the data space.

Regarding the effectiveness of the selected DR techniques, a total of twelve hypotheses are formulated before the study. They are derived based on the summarized advantages of the DR techniques from our literature review. In the experiment design, three key design choices are made to ensure the soundness of the experiment and obtain compelling results. First, we select a total of eight datasets whose dimension count, data points count, and cluster patterns are balanced in the design. Then, a pre-study is conducted to tune and optimize the parameters for the selected DR techniques, ensuring their robustness for each of the datasets. In addition to the four controlled experiments, a subjective experiment is conducted where the participants rank the clustering outcomes based on their personal preference. It allows us to evaluate the DR from a more subjective perspective. 

Based on the experiment results, we perform a comprehensive statistical analysis. The analysis results of the objective metrics validate seven of the hypotheses but leave five rejected. We found that although non-linear and local techniques show significant advantages in tasks such as cluster identification and cluster membership identification, there is not a technique that can outperform the others in all the four visual cluster tasks. 
Moreover, some linear techniques that receive less citation, such as NMF (Nonnegative Matrix Factorization)~\cite{lee1999learning} and t-SNLE (t-Distributed Stochastic Neighbor Linear Embedding)~\cite{bunte2012general}, catch our eyes because they yield better performance in tasks such as cluster distance comparison and density comparison, respectively. We also discuss the impact of the two DR properties on the effectiveness of the visual perception and the potential improvement of the experiment design. The DR techniques, datasets, and records of the study are available at \href{https://github.com/DR-approach/DR-approaches}{https://github.com/DR-approach/DR-approaches}.

\section{Related Work}


\subsection{DR techniques for visual cluster analysis}

Because our study focuses on the two essential properties of linearity and locality, we divide existing DR techniques into four groups as Espadoto et al.~\cite{espadoto2019towards}:
Non-Linear\&Local, Linear\&Local, Non-Linear\&Global, and Linear\&Global.

\textbf{Non-Linear\&Local} utilizes non-linear functions and preserves local neighborhood in DR processes. Representatives include LLE (Locally Linear Embedding)~\cite{roweis2000nonlinear}, LE (Laplacian Eigenmaps)~\cite{belkin2002laplacian}, LAMP (Local Affine Multidimensional Projection)~\cite{joia2011local}, t-SNE~\cite{maaten2008visualizing}, and UMAP~\cite{mcinnes2018umap}. They are usually considered as manifold learning techniques that can capture the clusters in manifolds.
For example, LLE preserves the linear combination of local neighborhoods. 
To achieve a similar goal, LE embeds data points according to the eigenfunctions of the Laplace Beltrami operator on the manifold. LAMP uses orthogonal mapping theory to build accurate local transformations. Especially, t-SNE and UMAP are widely used in visual cluster analysis because of their robust performance on clustering separation ~\cite{arora2018analysis, linderman2019clustering}.

\textbf{Linear\&Local} utilizes linear functions and preserves local neighborhood in DR processes. Representatives include LPP (Locality Preserving Projection)~\cite{he2004locality}, NPE (Neighborhood Preserving Embedding)~\cite{he2005neighborhood}, and t-SNLE~\cite{bunte2012general}.   
All the three techniques are linear variants of LE, LLE, and t-SNE, respectively. Compared to their non-linear counterparts, they retain the advantage in clustering separation but are limited by the linear projection~\cite{bunte2012general}.  

\textbf{Non-Linear\&Global} utilizes non-linear functions to preserve the pairwise distance among all the data points. Representatives include LSP (Least Square Projection)~\cite{paulovich2008least}, MDS (non-linear version)~\cite{kruskal1978multidimensional,ingram2008glimmer}, and Isomap~\cite{tenenbaum2000global}. LSP uses least squares approximations. The non-linear version of MDS preserves distances of high-dimensional space in the embedded low-dimensional space.
Different from MDS, Isomap attempts to preserve the geodesic distance between any two pairs of data points. 
The scheme enables a consistent overview of the data based on the globally optimal layouts ~\cite{cheng2015data}.

\textbf{Linear\&Global} uses linear functions to preserve the pairwise distance structure amongst all the data samples. Representatives include PCA~\cite{wold1987principal}, FA (Factor Analysis)~\cite{mair2018factor}, and NMF~\cite{lee1999learning}. PCA aims to preserve data patterns in the aspect of variance. Ding et al.~\cite{ding2004k} prove that principal components are actually the continuous solution of the cluster membership indicators in the K-means clustering method, which indicates that PCA implicitly performs clustering. Compared to PCA, FA intends to identify latent variables underlying a higher-dimensional space of measurements. NMF aims to find two non-negative matrices $W$ and $H$ whose product can well approximate the original matrix $V$, namely $W\times H \approx V$. The matrix $W$ is a low-dimensional representation of the original data~\cite{cai2008non}.



\subsection{Evaluations of DR techniques in the aspect of visual cluster/class analysis}
\textbf{Evaluation based on quantitative metrics}. A variety of quantitative metrics has been used to measure the spatial aspects of projections generated by different DR techniques. 
The metrics can be used either standalone to gauge desirable aspects of a projection, or jointly to assess the overall quality of outcomes. 
For example, aggregate metrics such as trustworthiness, continuity, neighborhood hit, distance and class consistency can be used to assess the quality of data clusters in the projected space~\cite{espadoto2019towards}. Local metrics, such as projection precision score~\cite{schreck2010techniques} and average local errors~\cite{martins2014visual}, separately compare small neighborhoods in a projection, upon which the detailed levels of the result can be assessed. To further assess the effectiveness of human perception on the projection, Aupetit et al.~\cite{aupetit2016sepme} proposed a set of visual measures that can mimic the human notion of separability in data classes. By validating the measures with a machine learning framework, they found the average proportion of same-class neighbors has the best prediction accuracy on human perception. Unlike the above measures~\cite{aupetit2016sepme}, which are specific to color-coded scatterplots,
Clumpy scagnostic~\cite{wilkinson2008scagnostics} and ClustMe~\cite{abbas2019clustme} are visual quality measures to quantify clusters in monochrome scatterplots. ClustMe was based on a quantitative study of cluster counting task. Wang et al.~\cite{wang2017perception} used GONG ~\cite{aupetit2002gamma} and SC (silhouette coefficient) ~\cite{rousseeuw1987silhouettes} to compare the class separation performance of 10 DR techniques such as PCA and t-SNE on 93 datasets. Spathis et al.~\cite{spathis2018fast} used SC to evaluate the performance of interactive DR techniques such as LAMP, PLMP (Part-Linear Multidimensional Projection), and KELP (Kernel-based Linear Projection). Vernier et al~\cite{vernier2020quantitative} evaluated performance of eleven DR techniques on ten datasets using the neighborhood preservation.







\textbf{Evaluation based on user perception}. Lewis et al. \cite{lewis2012human} and Sedlmair et al.~\cite{sedlmair2012taxonomy} reported that quantitative metrics might perform differently from human perception in cluster recognition and visual class separation tasks, respectively. A set of researches focus on perception based evaluations. Lewis et al.~\cite{lewis2012behavioral} designed a study to investigate human agreement on the embedding quality. Their results show that expert users are reasonably consistent judges of embedding quality, whereas novice users are very inconsistent with one another. Etemadpour et al.~\cite{etemadpour2014perception} conducted a controlled experiment to evaluate the performance of DR techniques on five visual cluster analysis tasks. Ventocilla et al.~\cite{ventocilla2020comparative} conducted a similar perception evaluation but with foci on PCA, t-SNE, Radviz, and SC. Xu et al.~\cite{xu2017evaluating} conducted an experiment to compare advanced feature transformation technology (FT-high) and traditional DR techniques, such as PCA and MCML, regarding their performance in visual cluster analysis tasks. The results show that FT-high would yield better results than the DR techniques, especially in the accuracy of identifying clusters. Zhao et al.~\cite{zhao2018evaluating} conducted a controlled experiment to evaluate the performance of four DR techniques on fuzzy clusters analysis. Nevertheless, none of the above experimental studies has considered the two important properties, namely linearity and locality, in terms of their impact on the performance of the DR techniques. Another important difference lies in the task design and measure. Most existing studies are based on color-encoded scatterplots rather than monochrome scatterplots. In the task of identifying clusters~\cite{etemadpour2014perception}, although monochrome scatterplots are used, their metric is based on the number of identified clusters only and lacks an accurate measure. Furthermore, most existing studies do not explore other cluster properties, such as density and membership.

\section{Evaluation Landscape}
In order to establish a solid basis for our study, we have conducted a comprehensive literature review, upon which the DR techniques, the datasets, and the analytical tasks of the study are derived. 
\subsection{Selection of DR techniques}
To select the most representative techniques from the wealth of existing DR research, we consider the following three metrics: (1) \textit{influence} - the techniques should be influential in general research fields, and more importantly, in the visualization field as well. In practice, we consider the citation as a major indicator of the influence; (2) \textit{task-suitability} - the techniques should have been widely applied in visual cluster analysis, and considered essential for accomplishing the related analytical tasks. Specifically, we focus on those that are the most cited techniques for certain visual cluster analysis tasks; (3) \textit{comparison-balance} - since we intend to investigate the techniques based on their high-level categories, each category should have a consistent number of techniques for an effective comparison.

Based on the first two metrics, we conducted a comprehensive literature survey on recent research papers in the visualization community. Specifically, the source includes papers published in the journal of IEEE TVCG and three mainstream visualization conferences (IEEE VIS, EuroVis, and PacificVis) between 2010 and 2020. We distilled 937 papers that not only consist of the keywords "cluster" or "clustering" but also are relevant to DR techniques. Among them, we narrowed the keyword search related to the visual cluster analysis, which derived 51 papers as the finally refined source. 

We conducted an in-depth survey on each of the papers and selected nine DR techniques based on our metrics. They were grouped based on the properties of linearity and locality. Table~\ref{tab:dr} summarizes the selected techniques with their categories, their google scholar citation counts, and the number of times it has been cited within the 51 papers. The classical techniques, such as PCA, have multiple related papers that have high citation counts. For clarity, we selected the paper with the highest citation counts for a technique. While there was no DR technique in the group of Linear\&Local, we selected three more techniques, namely LPP, NPE, and t-SNLE, which are linear variants of LE, LLE, and t-SNE, respectively. Similar to Espadoto et al.~\cite{espadoto2019towards}, these twelve techniques were distributed into four groups, including Linear \& Local (L\&Lc), Linear \& Global (L\&G), Non-linear\&Local (NL\&Lc), and Non-linear \& Global (N\&G) (see Table \ref{tab:dr}).

\begin{table}[!tb]      
\footnotesize
\centering
    \begin{tabular}{lllrr}    
        \toprule            
        Techniques & Linearity & Locality & Citation & NO. of Applications         \\
        \midrule            
        t-SNE~\cite{maaten2008visualizing} & non-linear & local & 18655 & 30 \\
        Isomap~\cite{tenenbaum2000global} & non-linear & global & 14394 & 2 \\
        UMAP~\cite{mcinnes2018umap} & non-linear & local & 2244 & 2\\
        MDS~\cite{kruskal1978multidimensional} & non-linear & global & 7942 & 19 \\
        LE~\cite{belkin2002laplacian} & non-linear & local & 4803 & 2 \\
        LLE~\cite{roweis2000nonlinear} & non-linear & local & 15811 & 2 \\
        PCA~\cite{wold1987principal} & linear & global & 8807 & 14 \\
        FA~\cite{mair2018factor} & linear & global & 14660 & 1\\
        NMF~\cite{lee1999learning} & linear & global & 12365 & 1 \\
        LPP~\cite{he2004locality} & linear & local & 4702 & 0 \\
        NPE~\cite{he2005neighborhood} & linear & local &  1829 & 0 \\
        t-SNLE~\cite{bunte2012general} & linear & local & 87 & 0 \\
        \bottomrule         
        \end{tabular}
        \renewcommand\tablename{Table}
        \caption{DR techniques used in our evaluation.}
        \label{tab:dr}
\end{table}

\begin{table}[!tb] 
\setlength{\belowcaptionskip}{-0.5cm} 
\footnotesize
\centering
    \begin{tabular}{lrrr}    
        \toprule            
      Datasets & Data Items & Dimensions & Clusters \\
        \midrule            
      EcoliProteins~\cite{Dua2019uci} & 336 & 7 & 8  \\
      Dermatology~\cite{Dua2019uci} & 259 & 34 & 6  \\
      ExtYaleB~\cite{georghiades2001few} & 320 & 30 & 5\\
      World12d~\cite{sedlmair2012taxonomy} & 151 & 12 & 5 \\
      Boston~\cite{sedlmair2012taxonomy} & 155 & 13 & 3 \\ 
      Mnist64~\cite{Dua2019uci} & 1083 & 64 & 6  \\
      Weather~\cite{ventocilla2020comparative} & 366 & 194 & 7  \\
      Olive~\cite{forina1983classification} & 572 & 8 & 3  \\
        \bottomrule         
        \end{tabular}
        \renewcommand\tablename{Table}
        \caption{Datasets used in our evaluation.}
        \label{tab:ds}
\end{table}

\subsection{Selection of Datasets}

To reflect the common interests of visualization community,
we select eight datasets out of 41 datasets that were used in the visualization applications in our survey. First, we considered the data size within a range from 100 to 1500 data instances and the cluster size from 3 to 10 to avoid scalability issue. Second, in order to make sure that the study can distinguish the performance of DR techniques, we select datasets for which cluster patterns can be visually separated in at least one projection result of a DR technique. Due to the huge decision space (41 datasets and 12 DR techniques), we assumed classes given by data labels formed clearly separable clusters. We used visual separation measures~\cite{aupetit2016sepme} GONG and KNNG to quantify class separation and select the best projections for our study. Noting that we have not fixed the DR technique that can visually separate clusters, selecting datasets with visually distinguishable clusters introduces few bias into the study, if there is. As a result, the selected eight datasets are shown in Table~\ref{tab:ds}).
We use PCA to reduce the dimensionality of ExtYaleB from 32,256  to 30 for a meaningful projection. All datasets are normalized by dimension.

\subsection{Selection of Typical Tasks}
The goal of the task formulation is to select the most common analytical tasks of the visual cluster analysis within general application domains. To this end, we extracted all the analytical tasks from the 51 surveyed cluster analysis papers and selected the top four most-cited tasks. 

\begin{itemize}
  \item\textbf{Cluster identification (T1)}: given a scatterplot generated by a DR technique, identify dense and well-separated clusters. Input: projected data. Output: color-coded points from each identified cluster. Action: lasso selection.
  \vspace{-2mm}
  \item \textbf{Membership identification (T2)}: given a point in a scatterplot, identify the cluster it belongs to. Input: projected data and one color-coded target point. Output: color-coded points of the cluster identified to contain the target point. Action: lasso selection.
  \vspace{-2mm}
  \item \textbf{Distance comparison (T3)}: given a cluster in a scatterplot, identify the nearest cluster to it. Input: projected data and color-coded  points from one target cluster. Output: color-coded points from the cluster identified to be nearest to the target one. Action: lasso selection.
  \vspace{-2mm}
  \item \textbf{Density comparison (T4)}: given multiple clusters, identify the cluster with the largest density. Input: projected data. Output: color-coded points from the cluster identified to be of largest density. Action: lasso selection.
\end{itemize}

All tasks are completed in projection space. For each task, we created an experiment to compare the performance of different DR techniques. They are described in Section \ref{sec:experiment}.

\section{Pre-Study: Parameters of DR techniques}

The projection of the DR technique is usually significantly affected by their hyperparameters~\cite{garcia2013stability}. Improper parameters will lead to unreliable comparison results. In order to optimize the parameters of selected DR techniques for each dataset, we have conducted a preliminary experiment.


\subsection{Parameters}
In this study, we focus on hyperparameters that significantly affect the projection results and require trial-and-error tuning. Specifically, we considered two parameters, namely \textit{neighborhood value} and \textit{perplexity}, as they significantly affect the projection quality of LPP, NPE, Isomap, UMAP, LE, LLE, t-SNE, and t-SNLE in different degrees~\cite{van2009dimensionality,espadoto2019towards,garcia2013stability}. For the other parameters, such as iteration times,  we adopted recommended values based on previous studies~\cite{maaten2008visualizing,sedlmair2012taxonomy}. 
Following the previous studies ~\cite{espadoto2019towards,van2009dimensionality}, we formulated 5-level ranges for the parameter \textit{neighborhood value} and \textit{perplexity}, from which the optimal values can be selected for each DR method. The parameter range formulation ensures a moderate degree of discrimination for the projection results while keeping a reasonable range to fit the optimal values. The settings are presented in Table~\ref{tab:psps}.

\begin{table}[tb]
\setlength{\belowcaptionskip}{-0.5cm} 
\footnotesize
 \centering
 \begin{tabular}{llrl}
  \toprule
  Techniques & Key Parameters & Settings &  Other Parameters\\
  \midrule
 t-SNE & \textit{Perplexity} & 5,15,30,40,50 & \makecell[l]{\textit{Early exaggeration}=6\\ \textit{Iteration}=3000} \\
t-SNLE & \textit{Perplexity} & 5,15,30,40,50& \makecell[l]{\textit{Early exaggeration}=6\\ \textit{Iteration}=3000} \\
  UMAP & \textit{Neighbors} & 4,7,10,13,16 & \makecell[l]{\textit{Initialization}=spectral\\ \textit{Mininum distance}=0.1\\ \textit{Iteration}=500} \\
  LLE & \textit{Neighbors} &4,7,10,13,16 & \makecell[l]{\textit{IReg}=0.001\\ \textit{Iteration}=200} \\
 LPP & \textit{Neighbors} & 4,7,10,13,16 & None \\
 NPE & \textit{Neighbors} & 4,7,10,13,16 & None \\
 Isomap & \textit{Neighbors} & 4,7,10,13,16 & None \\
 LE & \textit{Neighbors} & 4,7,10,13,16 & None \\
 \midrule
 MDS & None &  & \textit{Iteration}=500 \\
 FA & None &  & \textit{Iteration}=2000 \\
 NMF & None &  & \textit{Iteration}=400 \\
 PCA & None &  & No parameters \\

  \bottomrule
 \end{tabular}
 \renewcommand\tablename{Table}
 \caption{The parameter settings of DR techniques in the pre-study. The formal study used the same settings except for \textit{perplexity} and \textit{neighbors}.}
 \label{tab:psps}
\end{table}

\subsection{Task and Procedure}

The same set of datasets was used in both formal study and pre-study. For a given dataset and a DR technique, participants were provided with 5 projections resulting from 5 parameter settings (see Table \ref{tab:psps}). The number and size of classes, which are taken as ground truth for clusters, are also expicitly mentioned to the participants of the pre-study (see figure 5 in 7.1 of supplemental material). It is worth mentioning that this is only for the pre-study to select good projections, not for the formal study. They were asked to select the projection that best fits the description: \textbf{``Please select the scatterplot that presents the cluster structure of the given dataset most actually''}.
For each participant, a total of $8(DR\ techniques )\times8(datasets)=64$ trials were evaluated.  






\textbf{Participants}.
We recruited 20 participants (16 males, 4 females) who are graduate students majoring in computer science. None of the participants reported color blindness or color weakness. 

\textbf{Procedure}. The preliminary experiment was conducted on an online application that enables a selection of DR projections for the pre-processed datasets. Participants were instructed to log in to the application to complete the tasks remotely. The tasks were conducted on standard laptops with a 1,920×1,080 screen resolution using the Chrome browser. Before completing the tasks, each participant was provided with a 10-minute background introduction, including research purposes, related concepts, experimental procedures, and precautions. Then, the participants were instructed to finish a total of 64 trials in random order. There was no time limit for task completion, but all the participants finished within 50 minutes. To track the progress, the participants were asked to share the screen with a remote instructor during the experiments. 

\subsection{Results}

\begin{table}[tb]     
\setlength{\belowcaptionskip}{-0.5cm} 
\centering
\begin{tabular}{lrrrrrrrr} 
\toprule  
\ &1&2&3&4&5&6&7&8\\
\midrule            
t-SNE & 40 & 5 & 15 & 5 & 15 & 40 & 5 & 40 \\
t-SNLE & 30 & 15 & 30 & 30 & 15 & 40 & 5 & 40 \\
UMAP & 13 & 16 & 10 & 7 & 7 & 16 & 4 & 10 \\
Isomap & 16 & 4 & 10 & 4 & 10 & 10 & 13 & 10 \\
LLE & 4 & 10 & 10 & 4 & 10 & 10 & 4 & 7 \\
LE & 4 & 4  & 7 & 10 & 10 & 16 & 10 & 13 \\
LPP & 10 & 7 & 10 & 13 & 7 & 4 & 10 & 16 \\
NPE & 4 & 10 & 7 & 4 & 4 & 16 & 13 & 7 \\
\bottomrule         
\end{tabular}
\renewcommand\tablename{Table}
\caption{The optimal parameters, i.e., the \textit{perplexity} and \textit{neighbors} (see Table \ref{tab:psps}), of DR techniques on the 8 datasets. Each dataset is represented by a serial number ID from 1 to 8, which corresponds to the 8 datasets in an order of Weather, Dermatology, ExtYaleB, World12d, Boston, Mnist64, EcoliProteins, and Olive, respectively.}
\label{tab:best}
\end{table}

For a given DR technique and a dataset, we retrieved the projection with the majority vote and considered the corresponding parameter values optimal for that dataset. The result is shown in Table~\ref{tab:best}. Based on the results, we utilized the optimal values of the DR techniques for the corresponding dataset in the formal study. 


\section{Formal Study}

\subsection{Hypotheses}

This formal study aims to evaluate the performance of twelve selected DR techniques on four common visual cluster analysis tasks. Please refer to the tasks from section 3. Based on the general views from our literature review and practical experience, we put forward twelve hypotheses to verify the possible performance differences of different types of DR techniques on the four tasks.




\noindent\textbf{H1.1}--\textbf{H1.3} Regarding \textbf{T1}, existing research pointed out that local DR techniques often ensure a clear visual separation of data entities in projections~\cite{he2004locality,maaten2008visualizing}. It is also pointed out that non-linear DR techniques favor the optimization of neighborhood distances and mostly disregard large distances~\cite{chatzimparmpas2020t}. Again, this might have positive effects on differentiating clusters in the projections.
Moreover, most applications employed t-SNE and UMAP for cluster analysis. Therefore, we have the following hypotheses:

\begin{itemize}
\vspace{-0.3cm}\item\noindent\textbf{H1.1} Local DR techniques perform better than global ones in \textbf{T1} if they have the same linearity type;
\vspace{-0.3cm}\item\noindent\textbf{H1.2} Non-linear DR techniques perform better than linear ones in \textbf{T1} if they have the same locality type;
\vspace{-0.3cm}\item\noindent\textbf{H1.3} t-SNE and UMAP perform better than other techniques in \textbf{T1} .
\end{itemize}




\vspace{-0.3cm}\noindent\textbf{H2.1}--\textbf{H2.3} Regarding  \textbf{T2}, although it is a different perception task from \textbf{T1}, it also requires the ability to separate clusters and preserve near neighbors. Therefore, we have hypotheses \textbf{H2.1}, \textbf{H2.2}, and \textbf{H2.3} in \textbf{T2} that are identical
to \textbf{H1.1}, \textbf{H1.2}, and \textbf{H1.3} in \textbf{T1}, respectively. 







\noindent\textbf{H3.1}--\textbf{H3.3} Regarding \textbf{T3}, global DR techniques are designed to preserve global pairwise distances and thus could enable more faithful projections of global structures, e.g., cluster relationships, of high-dimensional spaces~\cite{jolliffe1986principal}. Compared to the non-linear embeddings, the linear embeddings are more reliable in terms of metric-preserving properties~\cite{koren2003visualization}. This might facilitate the representation of cluster relationships. Moreover, PCA is the most widely used global linear technique. Therefore, we have the following hypotheses:

\begin{itemize}
\vspace{-0.3cm}\item\noindent\textbf{H3.1} Global DR techniques perform better than local ones in \textbf{T3} when they have the same linearity;

\vspace{-0.3cm}\item\noindent\textbf{H3.2} Linear DR techniques perform better than non-linear ones in \textbf{T3} when they have the same locality;

\vspace{-0.3cm}\item\noindent\textbf{H3.3} PCA performs better than other techniques in \textbf{T3}.
\end{itemize}




\vspace{-0.3cm}\noindent\textbf{H4.1}--\textbf{H4.3} When considering \textbf{T4}, there are many practical experiences~\cite{wattenberg2016use, narayan2021assessing} pointing out that non-linear or local techniques neglect the density information. It is also considered that PCA performs better than non-linear techniques~\cite{chatzimparmpas2020t}. Therefore, we have hypotheses \textbf{H4.1}, \textbf{H4.2}, and \textbf{H4.3} in \textbf{T2} that are identical to \textbf{H3.1}, \textbf{H3.2}, and \textbf{H3.3} in \textbf{T3}, respectively. 




\subsection{Experiments} \label{sec:experiment}
Under the guidance of the four tasks (\textbf{T1-T4}) and the associated twelve hypotheses (\textbf{H1.1-H4.3}), we designed four controlled experiments: \textbf{E1}--\textbf{E4} for \textbf{T1}--\textbf{T4} to verify \textbf{H1}--\textbf{H4}, respectively. In addition to the four controlled experiments, a subjective evaluation \textbf{E5} was designed where we studied participants' preferences for different DR techniques based on their free exploration of datasets. All DR techniques and datasets for \textbf{E1}--\textbf{E5} are the same. There were $12(techniques)\times8(datasets)=96$ trials for each participant in \textbf{E1}--\textbf{E4} and $1(3\times4 \ layout \  for \ 12 \ techniques)\times8(datasets)=8$ trials for each participant in \textbf{E5}.

\textbf{E1:} \textbf{T1} was performed to test if projected clusters were separated cleanly and easily recognized by participants. In each trial, the participants were asked to \textbf{use a lasso tool to interactively select well-separated clusters, if any, from the visualization generated by a DR technique.} Compared with other evaluation methods such as counting the cluster number ~\cite{etemadpour2014perception, abbas2019clustme}, interactive selections enable more accurate reflection of participants' visual perceptions, especially for handling aggregated clusters that might lead to bias in counting the cluster number~\cite{aupetit2014sanity}. Since color would interfere with the judgments of the clusters, we rendered all data points in grey color and full opacity. 


\textbf{E2:} \textbf{T2} was performed to test whether a meaningful belonging relationship between data points and clusters can be easily identified. In each trial, a pre-defined data point was highlighted in red color. Participants were asked to \textbf{use a lasso tool to interactively select a well-separated cluster, if any, that contains the highlighted data point.}
When selecting the pre-defined data point, we prefer challenging ones that are close to the cluster boundary in the high-dimensional space.
To this end, we extend the clustering metric of silhouette coefficient ~\cite{rousseeuw1987silhouettes} to high-dimensional space, upon which the silhouette coefficient values of each data point are calculated.
Then, the point with the lowest value was selected. 




\begin{figure}[tb]
\centering
\includegraphics[width=1\linewidth]{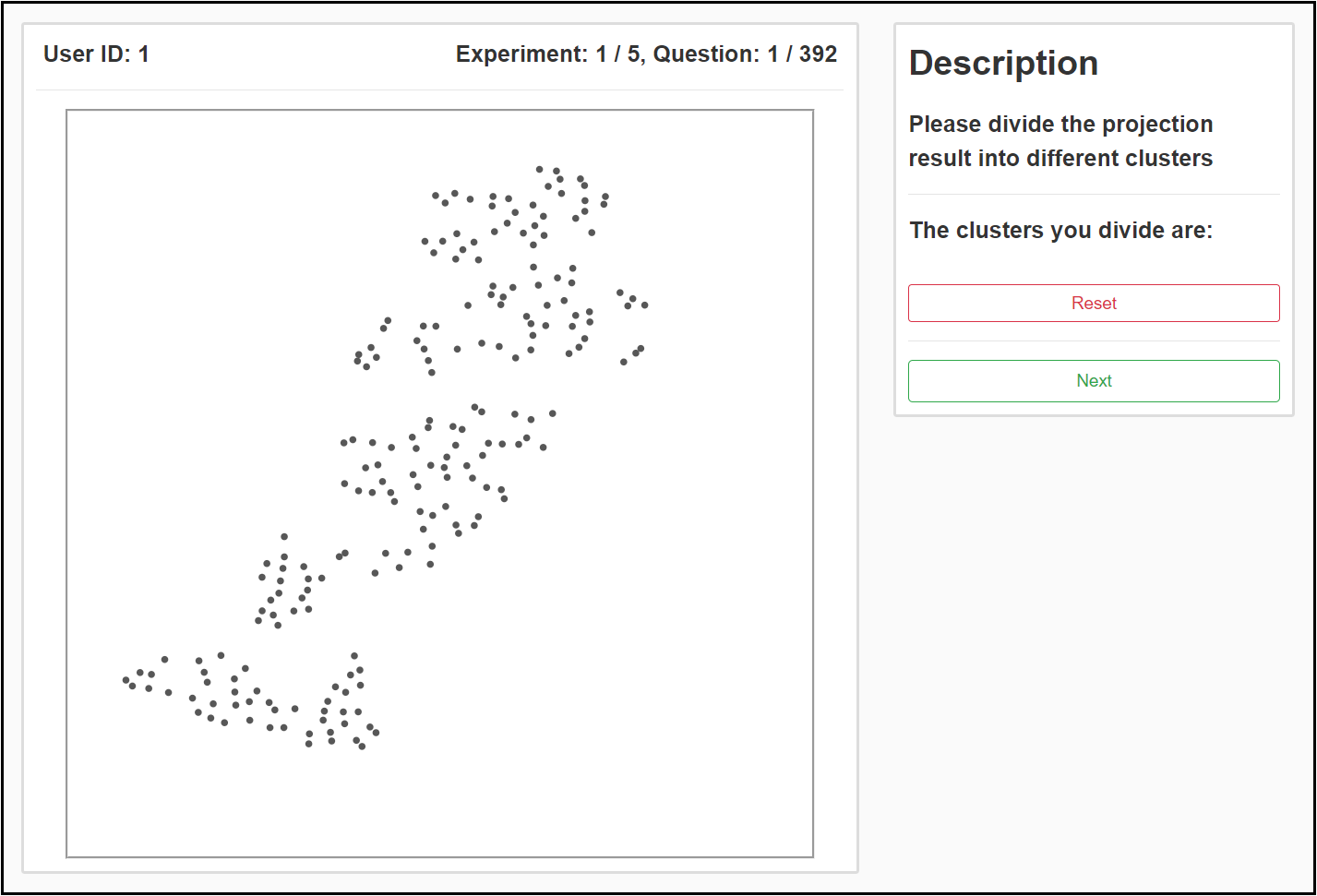}
\setlength{\abovecaptionskip}{0.cm}
\caption{Example interface of \textbf{E1} in the formal study. Participants were asked to identify clusters by lasso tool.}
\label{fig:web1} 
\end{figure}

\textbf{E3:} \textbf{T3} was performed to evaluate the ability of different DR techniques to preserve the relative distance among clusters in terms of visual perception. In each trial, a pre-defined cluster was highlighted in red color. Participants were asked to \textbf{use a lasso tool to interactively select a well-separated cluster, if any, that is closest to the given cluster.} 



\textbf{E4:} \textbf{T4} was performed to evaluate the ability of different DR techniques to preserve the density in the aspect of visual perception. In each trial, participants were asked to \textbf{use a lasso tool to interactively select a well-separated cluster, if any, that has the highest density among all clusters.} 





\begin{figure}[tb]
\setlength{\belowcaptionskip}{-0.5cm} 
\centering
\includegraphics[width=1\linewidth]{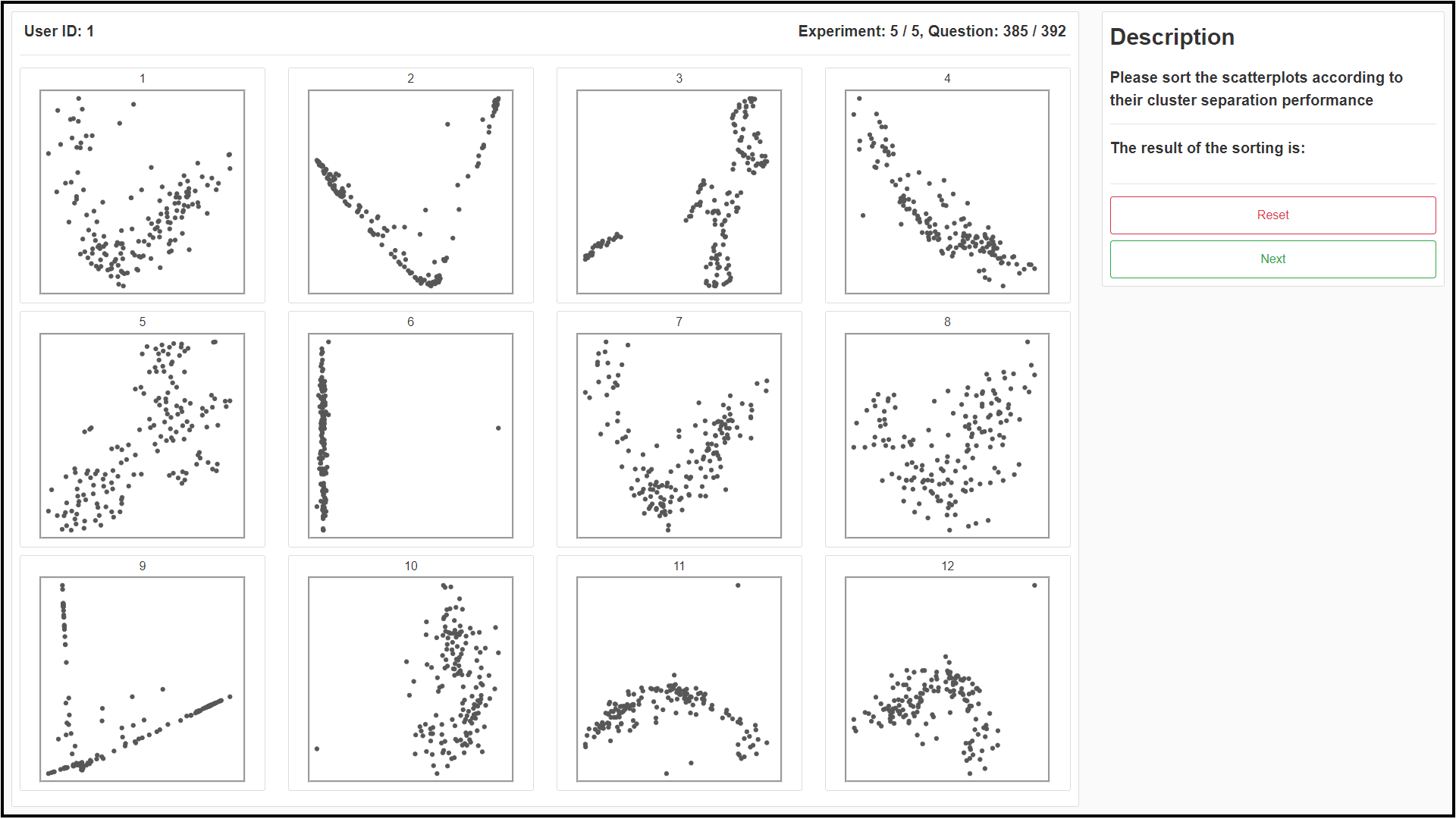}
\setlength{\abovecaptionskip}{0.cm}
\caption{Example interface of \textbf{E5} in the formal study. Participants were asked to rank the twelve visualizations according to their cluster separation performance.}
\label{fig:web} 
\end{figure}

\textbf{E5: Subjective experiment} was used to compare the overall performance of different DR techniques based on participants' personal preferences. In each trial, we assigned a participant twelve visualizations of a dataset generated by twelve DR techniques, respectively. They were asked to \textbf{rank the rank the visualization from 1 to 12 (1 is the best) by their personal preference in terms of cluster separation performance, identical ranking for multiple visualizations is allowed in case their qualities are hard to distinguish.} The visualizations were numbered and placed in a $3\times4$ layout with a random order (see Fig.\ref{fig:web}).

\subsection{Participants, Apparatus, and Testing Data}
\textbf{Participants}. We recruited 60 participants (40 males and 20 females) for the formal study. Their age was from 19 to 32 with an average of 24. 56 of them are graduate students and four of them are bachelor students with research experience. They come from majors including computer science (46), engineering (9), art (4), and mathematics (1). 19 participants have a data visualization background.
6 participants study clustering algorithms. 11 participants have performed cluster analysis in bioinformatics. Each participant was rewarded \$20 per hour for completing the experiments. None of them reported color blindness or color weakness. Participants in the formal study are different from participants in the pre-study. 


\noindent\textbf{Apparatus}. 
The formal study was conducted online on a pre-built web application in order to involve more participants. The application allows participants to complete all the controlled experiments and subjective studies (\textbf{E1}--\textbf{E5}), providing functionality such as visualizations, interactions, and time counter to facilitate the experiments. All the participants remotely took the experiments on standard laptops with a 1,920×1,080 screen resolution and chrome browser. The participants were asked to share their screens with the researcher during the experiments to enable remote monitoring.

\noindent\textbf{Testing Data}. Before the formal study, we generated scatterplots based on the eight selected datasets. For each dataset, we created twelve scatterplots by each of the selected DR techniques, respectively.
The parameter of each DR technique was determined based on the results of the pre-study.
The same layout was used for a specific DR technique and a dataset for different experiments. 
Except for the given points in \textbf{E2} and the given clusters in \textbf{E3}, which were coded in red, the data points in the scatter plots were all coded in gray. 
The points were rendered with a radius of 3 pixels without
transparency.
The size of the scatterplots was $600\times550$ pixels in \textbf{E1}--\textbf{E4}, and $300\times300$ pixels in \textbf{E1}.





\subsection{Procedure}
A training session was conducted before the formal experiments. For the training session, we utilized the Iris dataset~\cite{Dua2019uci} to generate scatterplots for each DR technique. In the training session, the instructor explained the related concepts and the experimental procedure. Then, the participant was asked to complete a total of thirteen training trials, including three trials for each controlled experiment (\textbf{E1}--\textbf{E4}) and one trial for the subjective experiment (\textbf{E5}). The DR technique of the trial was randomly selected for each experiment. After the training session, the participant was asked to complete \textbf{E1}--\textbf{E5} in order. A 5-minutes break was allowed before each experiment. The completion order of the trials in each experiment was randomly assigned to each participant. The participants used the application to complete each trial, upon which the trial completion time and results were automatically stored by the application. There was no time limit for each experiment. All the participants finished the five experiments within 120 minutes.  

Toward the end of \textbf{E5}, the participants were asked to complete a questionnaire about their background and subjective feedback of the experiments. The questionnaire includes three sections. The first section is about the participant background, such as the education level, visualization experience, and knowledge about the DR techniques. In the second section, the participant was asked to list the factors, including variance of density, cluster separation, variance of count, variance of size and variance of shape, if any, they relied on identifying the clusters in \textbf{E1}. In the last section, the participant was asked to rate the importance of the four analytical tasks (\textbf{T1-4}) in their data analysis experience using a five-point Likert scale. After the questionnaire, we conducted a subjective interview to understand how the scatterplots could affect participants' trial completion and their preference and/or experience on any DR techniques and visual clustering techniques in daily work.

\section{Statistical Analysis}


\begin{figure*}[!tb]
\centering
\includegraphics[width=0.97\linewidth]{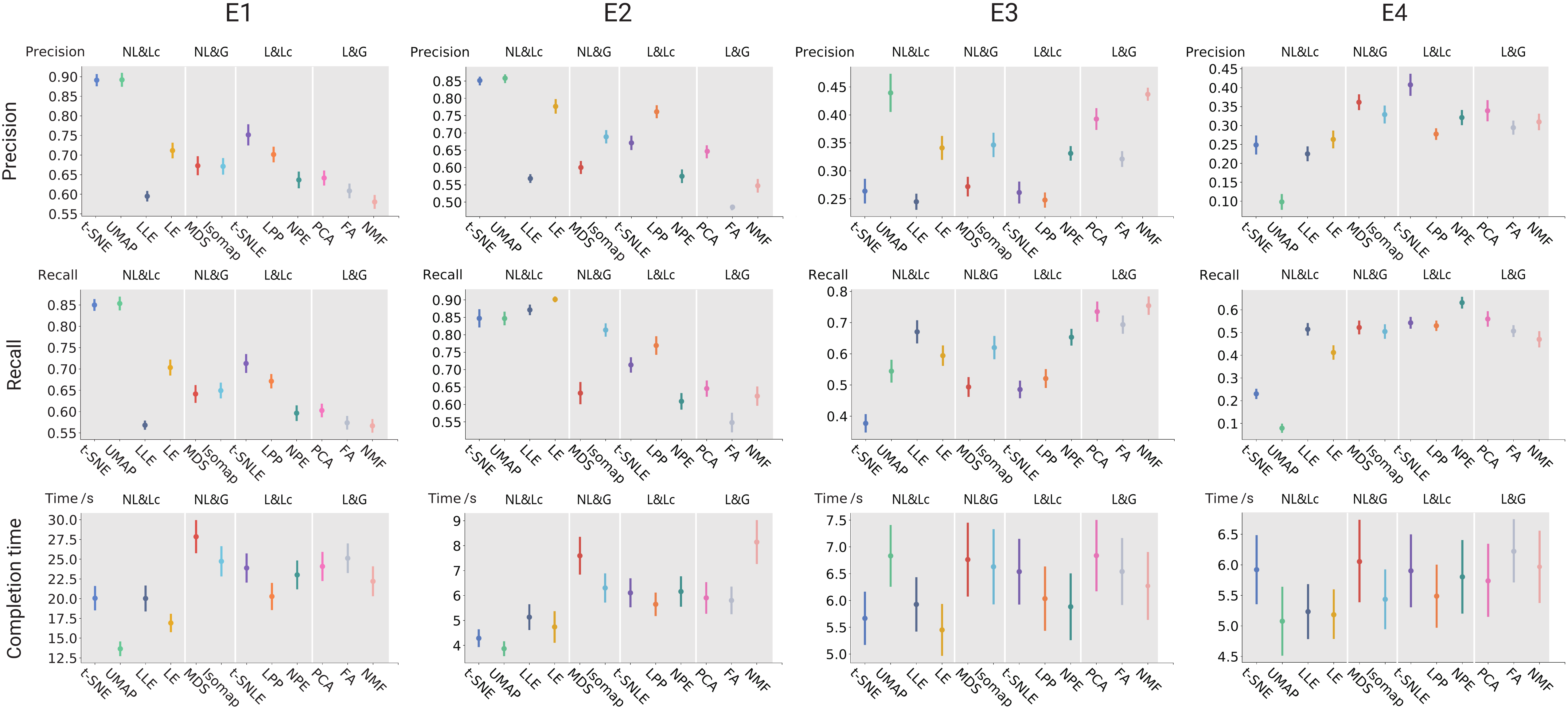}
\caption{Precision, Recall and Completion time of \textbf{E1--E4}. Error bars indicate 95\% confidence intervals.
}
\label{fig:errorbar} 
\end{figure*}

\subsection{Result Measurement}
To measure results of each trial for the five experiments, two conditions were considered: the trial completion time (the shorter the better), precision and recall (1 is best) were calculated for \textbf{E1}--\textbf{E4}; and the average ranking score of DR techniques was calculated for \textbf{E5}. The ground truth for computing the precision and recall are based on the class label of the datasets.

\noindent\textbf{E1}: While participants would identify different cluster structures, e.g., number of clusters, from the ground truth, matching the identification and the ground truth should be performed before the computation of precision and recall.
We denote the identified clusters by participants as $\{C_i\}_{i=1}^{K}$, where $K$ is the number of identified clusters. The ground-truth clusters are denoted as $\{G_j\}_{j=1}^{N}$, where $N$ is the number of labels in the dataset.
A matching function $f$ matches an identified cluster $C_i$ to $f(C_i)$, where $f(C_i) \in \{G_j\}_{j=1}^{N}$. 
Note that $f(C_i)$ could be an empty set when there is no matching for $C_i$. 
The match is based on two rules.
First, if an identified cluster $C_i$ contains data points from multiple ground-truth clusters, it is matched with the ground-truth cluster contributing the most data points to $C_i$.
Second, if multiple identified clusters are matched to the same ground-truth cluster following the first rule, only the identified cluster containing the most data points from the ground-truth cluster is accepted.
In each trial, we compute the precision as
$\frac{\sum_{i=1}^K{\vert C_i \cap f(C_i) \vert}}{\sum_{i=1}^K{\vert C_i \vert}}$, where $\vert C_i \vert$ is the size of $C_i$.
The recall is computed as 
$\frac{\sum_{i=1}^K{\vert C_i \cap f(C_i) \vert}}{\sum_{j=1}^N{\vert G_j \vert}}$.


\noindent\textbf{E2}--\textbf{E4}: 
The computation of precision and recall in \textbf{E2}--\textbf{E4} is simpler than \textbf{E1}, because there is only one identified cluster $C_i$.
We denote the ground truth cluster as $G_j$,
In each trial, the precision is the ratio of $\vert C_i \cap G_j \vert$ to $\vert C_i \vert$, and the recall is the ratio of $\vert C_i \cap G_j \vert$ to $\vert G_j \vert$. In \textbf{E3}, We calculated the distance among clusters in the high-dimensional space for the ground truth.
As Etemadpour et al.~\cite{etemadpour2014perception}, for each pair of clusters, we calculated the average pairwise Euclidean distances of all pairs of points belonging to the two clusters, respectively.
Based on the distance among clusters, we randomly selected one cluster as the pre-defined cluster and its nearest cluster as the ground truth. In \textbf{E4}, the participants' selection was tested against the ground truth that has the highest density based on the minimum spanning tree-based measure in the high-dimensional space~\cite{etemadpour2014perception}.

\noindent\textbf{E5}: For the subjective experiment \textbf{E5}, we compare the average ranking score of DR techniques in all trials. In each trial, the first one gets 12 points, the second one gets 11, ..., and the last one gets 1 point.

\begin{wrapfigure}{r}{4.5cm}
\vspace{-10pt}
\hspace{-10pt}
\centering
\includegraphics[width=0.25\textwidth]{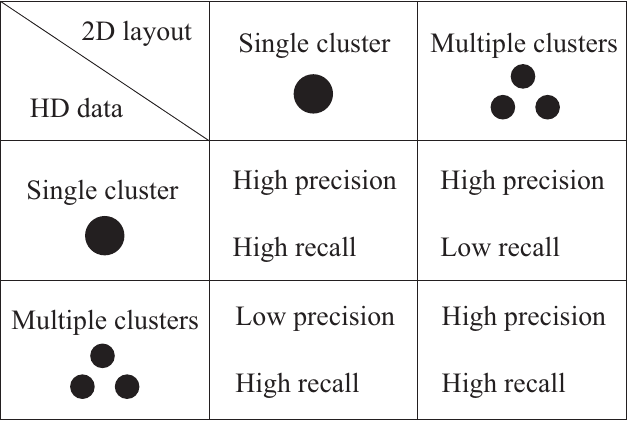}
\vspace{-5pt}
\hspace{-5pt}
\vspace{-5pt}
\label{fig:precision} 
\end{wrapfigure}

In \textbf{E1}--\textbf{E4}, For evaluating the performance of a DR technique in an experiment, we calculate the average precision and recall values on all participants and on all corresponding trials of the experiment. It is worth noting that both precision and recall are important for accurate measure of the quality. On one hand, when participants mistakenly identify unrelated points to the target cluster, the precision decreases.
On the other hand, if they correctly identify partial target cluster, the recall decreases.
As shown in the confusion matrix, a DR technique with high precision is capable of well separating ground truth clusters.
Otherwise, multiple ground truth clusters that mixed in the projection would result in low precision.
A low recall indicates that the DR technique cannot gather points together in one ground truth cluster, leading to a peak splitting event~\cite{xiang2019interactive}.
If a DR technique projects multiple ground truth clusters as one cluster in the 2D layout, participants would select all the connection points as one cluster, resulting in low precision and high recall.



\subsection{Statistical Analysis Approaches}
We utilize the average value and confidence interval of the result measures as indicators to test our hypotheses. Based on a Shapiro-Wilk test, we found that the results of \textbf{E1}--\textbf{E4} do not follow a normal distribution. Therefore, we employed a Friedman test to examine any significant differences among the various indicators for the DR techniques. For any resulted significant differences from the test, we further conducted detailed comparisons based on Tukey's HSD test. Here, we consider the standard significance level $\alpha$ = 0.05 as a statistical significance for the results.
When we take the effect size as 0.25 and the power as 0.95, the power analysis show that we need at least 18 samples in each experiment~\cite{prajapati2010sample}. In our experiment, we have 60 samples for each trial that is sufficient for the p-value test.


\subsection{Result Analysis}

\subsubsection{Objective Experiment Results}

\begin{figure}[!tb]
\centering
\includegraphics[width=1\linewidth]{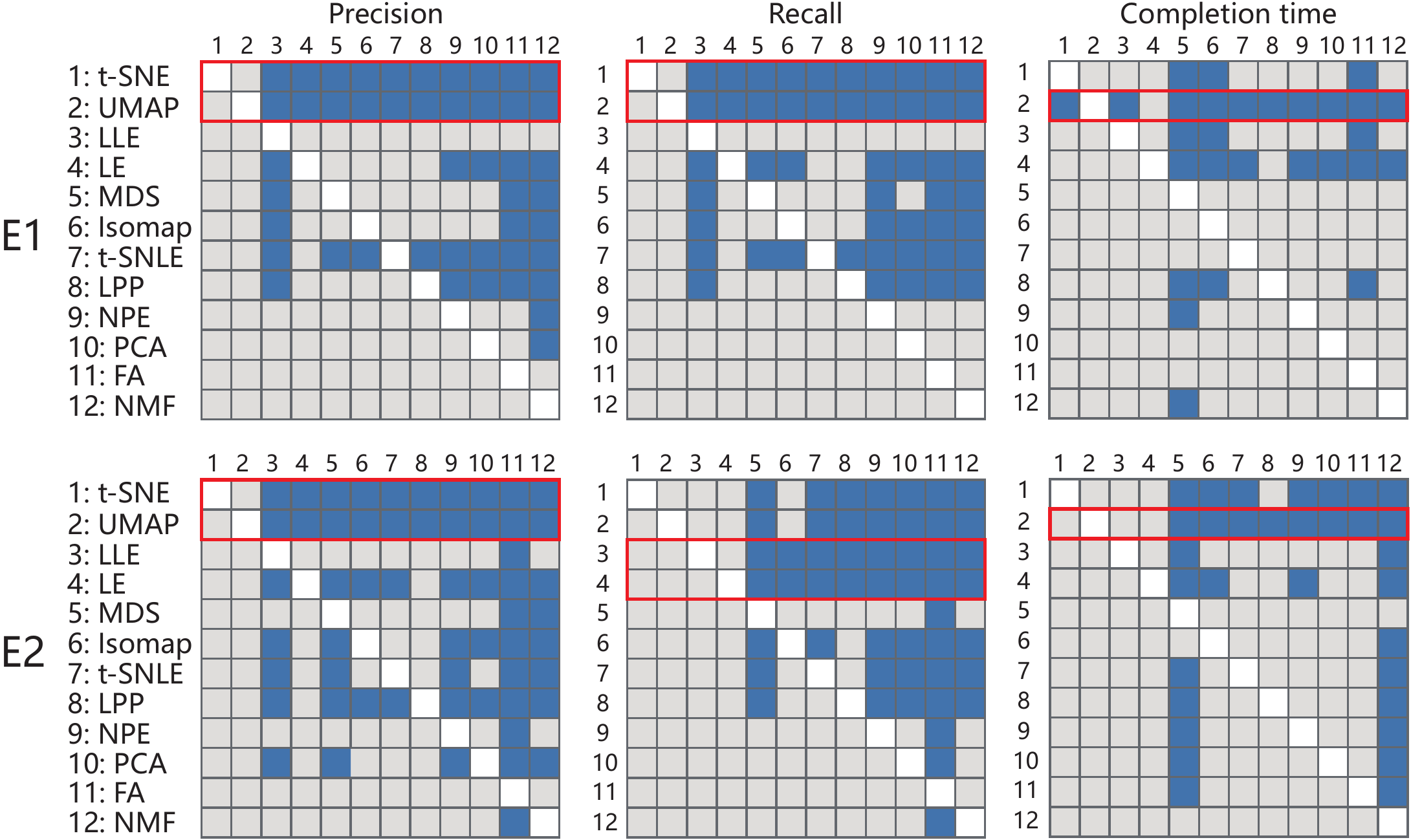}
\setlength{\abovecaptionskip}{0.cm}
 \caption{Matrix depiction of the pairwise significance relationships of the precision, recall and completion time differences of the projection techniques in \textbf{E1--E2}. A blue cell in row $i$ and column $j$ denotes that the technique with ID $i$ performs significantly better than the technique with ID $j$. The red rectangles denote the best DR techniques.}
\label{fig:hsd1} 
\end{figure}

\textbf{H1.1}--\textbf{H1.3} The results of \textbf{E1} are displayed in Fig.~\ref{fig:errorbar}.
Among all techniques, UMAP and t-SNE have the best precision (89.07\% and 85.34\%, respectively) and recall (88.98\% and 84.99\%, respectively). UMAP also has the shortest completion time (13.65s). The Friedman tests show 
statistical significance of precision ($\chi^{2}(11)$=524.90, $p$\textless0.05), recall($\chi^{2}(11)$=524.89, $p$\textless0.05) and average completion time ($\chi^{2}(11)$=261.82, $p$\textless0.05). Fig.~\ref{fig:hsd1} shows the pairwise significance relationships for all the DR techniques.


\begin{itemize}
 \vspace{-0.2cm}\item \noindent\textbf{H1.1} Local DR techniques perform better than global ones in \textbf{T1} if they have the same linearity type. \textbf{H1.1} is partially confirmed.
\end{itemize}

\vspace{-0.2cm}\noindent In the non-linear group (NL\&Lc and NL\&G), local techniques t-SNE and UMAP perform significantly better than all global techniques in terms of precision and recall. 
However, the local technique LLE performs significantly worse than global techniques MDS and Isomap. In the linear group (L\&Lc and L\&G), the local techniques t-SNLE and LPP performs significantly better than all the global techniques in terms of precision and recall. In terms of completion time, local techniques UMAP and LE perform significantly better than all global techniques.

\begin{itemize}
 \vspace{-0.2cm}\item\noindent\textbf{H1.2} Non-linear DR techniques perform better than linear ones in \textbf{T1} if they have the same locality type. \textbf{H1.2} is partially confirmed.
\end{itemize}

\vspace{-0.2cm}\noindent In terms of precision and recall, in the local group (NL\&Lc and L\&Lc), non-linear techniques t-SNE and UMAP perform significantly better than all linear techniques. However, the non-linear technique LLE performs significantly poorer than linear techniques t-SNLE and LPP. In the global group (NL\&G and L\&G), non-linear techniques MDS and Isomap perform significantly better than all the linear techniques except for PCA.

\begin{itemize}
\vspace{-0.2cm}\item\noindent\textbf{H1.3} t-SNE and UMAP perform better than other techniques in \textbf{T1}. \textbf{H1.3} is confirmed.
\end{itemize}

\vspace{-0.2cm}\noindent In \textbf{E1}, UMAP and t-SNE perform significantly better than all other techniques in precision and recall. They also perform well in completion time. In terms of completion time, UMAP performs significantly better than other techniques except LE.

\noindent\textbf{H2.1}--\textbf{H2.3} As shown in Fig.~\ref{fig:errorbar}, UMAP has the highest precision (85.81\%) and shortest completion time (3.87s) in
\textbf{E2}. t-SNE is the second in precision(85.14\%) and completion time (4.29s). The Friedman tests show significance among different DR techniques in terms of precision ($\chi^{2}(11)$=534.79,$p$\textless0.05), recall ($\chi^{2}(11)$=506.43,$p$\textless0.05) and average completion time($\chi^{2}(11)$=297.76,$p$\textless0.05). The pairwise significance relationships between each pair of DR techniques can be seen in Fig.~\ref{fig:hsd1}.

\begin{itemize}
 \vspace{-0.2cm}\item\noindent\textbf{H2.1} Local DR techniques perform better than global ones in \textbf{T2} if they have the same linearity type. \textbf{H2.1} is partially confirmed.
\end{itemize}

\vspace{-0.2cm}\noindent In the non-linear group (NL\&Lc and NL\&G), local techniques t-SNE and UMAP perform significantly better than all global techniques in terms of precision. However, the local technique LLE performs significantly poorer than the global technique Isomap. In terms of recall, local techniques perform significantly better than all global techniques except Isomap. In the linear group (L\&Lc and L\&G), in terms of precision and recall, local techniques t-SNLE and LPP perform significantly better than all global techniques except PCA. In terms of completion time, local techniques t-SNE and UMAP also perform significantly better than all global techniques.

\begin{itemize}
 \vspace{-0.2cm}\item\noindent\textbf{H2.2} Non-linear DR techniques perform better than linear ones in \textbf{T2} if they have the same locality. \textbf{H2.2} is partially confirmed.
\end{itemize}

\vspace{-0.2cm}\noindent In the local group (NL\&Lc and L\&lc), non-linear techniques t-SNE and UMAP perform significantly better than all linear techniques in terms of precision and recall. However, the non-linear technique LLE performs significantly poorer than the linear technique t-SNLE in precision. In the global group (NL\&G and L\&G), the non-linear technique Isomap performs significantly better than all linear techniques. In terms of completion time, non-linear techniques t-SNE and UMAP perform significantly better than all techniques except LPP. 

\begin{itemize}
 \vspace{-0.2cm}\item\noindent\textbf{H2.3} t-SNE and UMAP perform better than other techniques in \textbf{T2}. \textbf{H2.3} is partially confirmed.
\end{itemize}

 \vspace{-0.2cm}\noindent In \textbf{E2}, UMAP and t-SNE perform significantly better than all linear techniques in precision. Though they have lower recall than LE and LLE, there is no significant difference between the four techniques. In terms of completion time, UMAP and t-SNE also perform significantly better than the other techniques except LLE, LE and LPP.

\begin{figure}[!tb]
\centering
\includegraphics[width=1\linewidth]{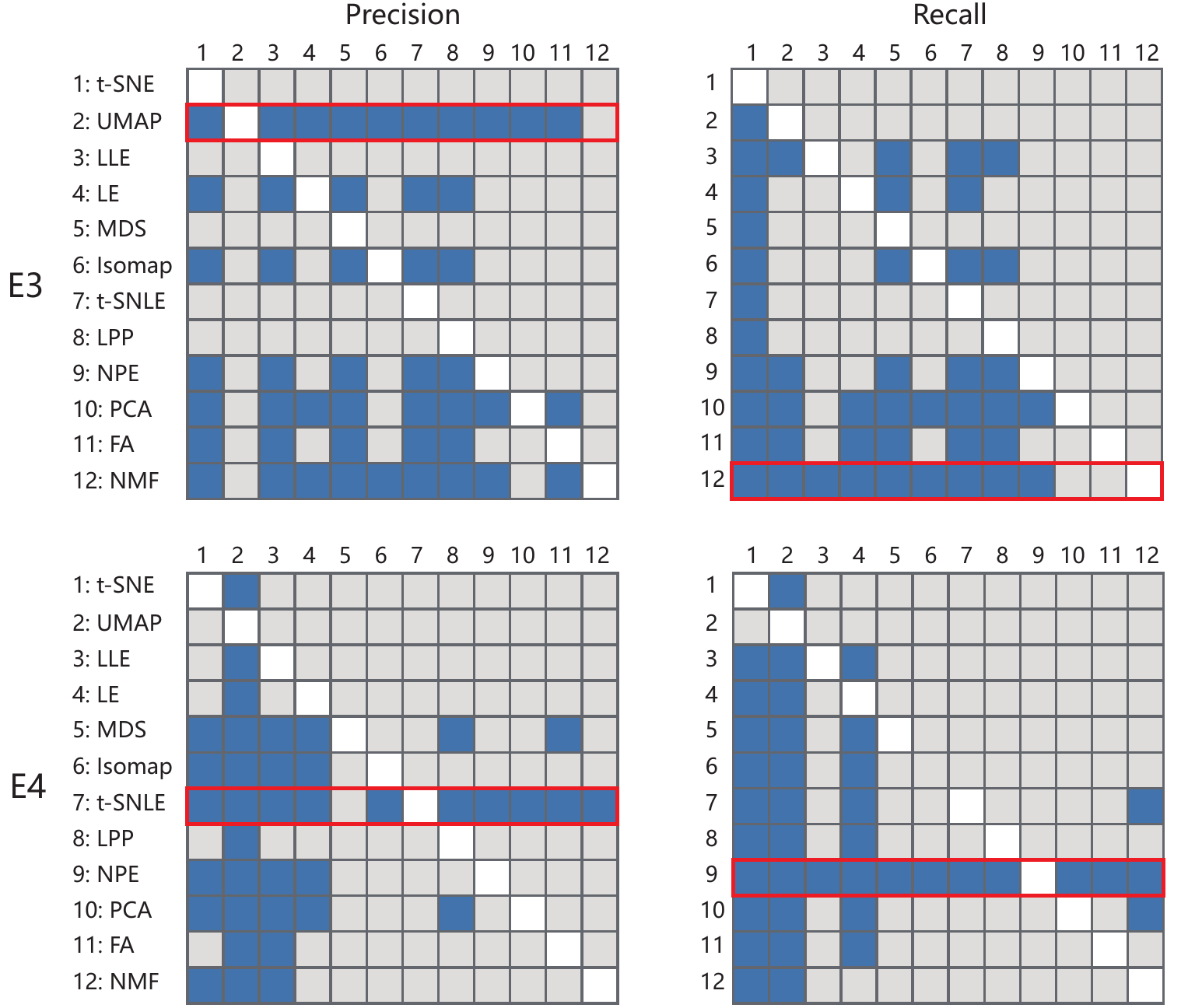}
\setlength{\abovecaptionskip}{0.cm}
\caption{Matrix depiction of the pairwise significance relationships of the precision and recall differences of the projection techniques in \textbf{E3--E4}. A blue cell in row $i$ and column $j$ denotes that the technique with ID $i$ performs significantly better than the technique with ID $j$. The red rectangles denote the best DR techniques.
}
\label{fig:hsd2} 
\end{figure}

\noindent\textbf{H3.1}--\textbf{H3.3} The results of \textbf{E3} are displayed in Fig.~\ref{fig:errorbar}. UMAP has the highest precision (43.94\%) but
ranked 8th in recall(54.42\%). NMF has the second-highest precision(43.69\%) and highest recall(75.42\%). The Friedman tests show statistical significance among different DR techniques in terms of precision ($\chi^{2}(11)$=534.79,$p$\textless0.05) and recall ($\chi^{2}(11)$=506.43,$p$\textless0.05). However, Tukey's HSD test shows there are no pairwise significance relationships exist between
each pair of DR techniques in terms of completion time. The pairwise significance relationships between each pair of DR techniques can be seen in Fig.~\ref{fig:hsd2}.

\begin{itemize}
 \vspace{-0.2cm}\item\noindent\textbf{H3.1} Global DR techniques perform better than local ones in \textbf{T3} when they have the same linearity. \textbf{H3.1} is rejected.
\end{itemize}

\vspace{-0.2cm}\noindent In the non-linear group (NL\&Lc and NL\&G), the local technique UMAP performs significantly better than all global techniques in terms of precision. 
In terms of recall, global techniques MDS and Isomap only significantly better than local techniques t-SNE. In the linear group (L\&Lc and L\&G), all global techniques perform significantly better than all local techniques except NPE in terms of precision and recall. 

\begin{itemize}
 \vspace{-0.2cm}\item\noindent\textbf{H3.2} Linear DR techniques perform better than non-linear ones in \textbf{T3} when they have the same locality. \textbf{H3.2} is rejected.
\end{itemize}

\vspace{-0.2cm}\noindent In the local group (NL\&Lc and L\&Lc), the non-linear technique UMAP performs significantly better than all linear techniques in terms of precision. The linear technique NPE performs significantly better than non-linear techniques t-SNE and LLE in precision. In the global group (NL\&G and L\&G), linear techniques PCA, FA, and NMF perform significantly better than the non-linear technique MDS in terms of precision and recall.

\begin{itemize}
 \vspace{-0.2cm}\item\noindent\textbf{H3.3} PCA performs better than other techniques in \textbf{T3}. \textbf{H3.3} is rejected.
\end{itemize}

\vspace{-0.2cm}\noindent In \textbf{E3}, PCA ranks as the third technique in precision (39.27\%) and the second technique in recall (73.5\%). UMAP performs significantly better than PCA in precision. 

\noindent\textbf{H4.1}--\textbf{H4.3} As shown in Fig.~\ref{fig:errorbar}, t-SNLE has the highest precision (40.77\%) and the third recall(54.29\%) in \textbf{E4}. NPE has the fifth precision(32.11\%) but the highest recall(63.15\%). The Friedman tests show statistical significance among different DR techniques in terms of precision ($\chi^{2}(11)$=247.55,$p$\textless0.05) and recall ($\chi^{2}(11)$=352.34.43,$p$\textless0.05). However, Tukey's HSD test shows there are no pairwise significance relationships exist between
each pair of DR techniques in terms of completion time. The pairwise significance relationships between each pair of DR techniques can be seen in Fig.~\ref{fig:hsd2}.

\begin{itemize}
 \vspace{-0.2cm}\item\noindent\textbf{H4.1} Global techniques perform better than local ones in \textbf{T4} if they have the same linearity. \textbf{H4.1} is rejected.
\end{itemize}

\vspace{-0.2cm}\noindent In the non-linear group (NL\&Lc and NL\&G), global techniques MDS and Isomap perform significantly better than all local techniques in terms of precision and recall. In the linear group (L\&Lc and L\&G), the local technique t-SNLE performs significantly better than all global techniques in terms of precision. In terms of recall, the local technique NPE performs significantly better than all global techniques. 

\begin{itemize}
 \vspace{-0.2cm}\item\noindent\textbf{H4.2} Linear techniques perform better than non-linear ones in \textbf{T4} if they have the same locality. \textbf{H4.2} is partially confirmed.
\end{itemize}

\vspace{-0.2cm}\noindent For the local group (NL\&Lc and L\&lc), in terms of precision, linear techniques t-SNLE and NPE perform significantly better than all non-linear techniques. In terms of recall, t-SNLE, LPP, and NPE also perform significantly better than all non-linear techniques except LLE. However, in the global group (NL\&G and L\&G), the non-linear technique MDS performs significantly better than the linear technique FA in terms of precision.

\begin{itemize}
 \vspace{-0.2cm}\item\noindent\textbf{H4.3} PCA performs better than other techniques in \textbf{T4}. \textbf{H4.3} is rejected.
\end{itemize}

\vspace{-0.2cm}\noindent In \textbf{E4}, PCA ranks as the third technique in precision (33.91\%) and the second technique in recall (55.96\%). t-SNLE and NPE perform significantly better than PCA in precision and recall, respectively.

\subsubsection{Subjective Experiment Results}

\begin{figure}[!tb]
\centering
\includegraphics[width=1\linewidth]{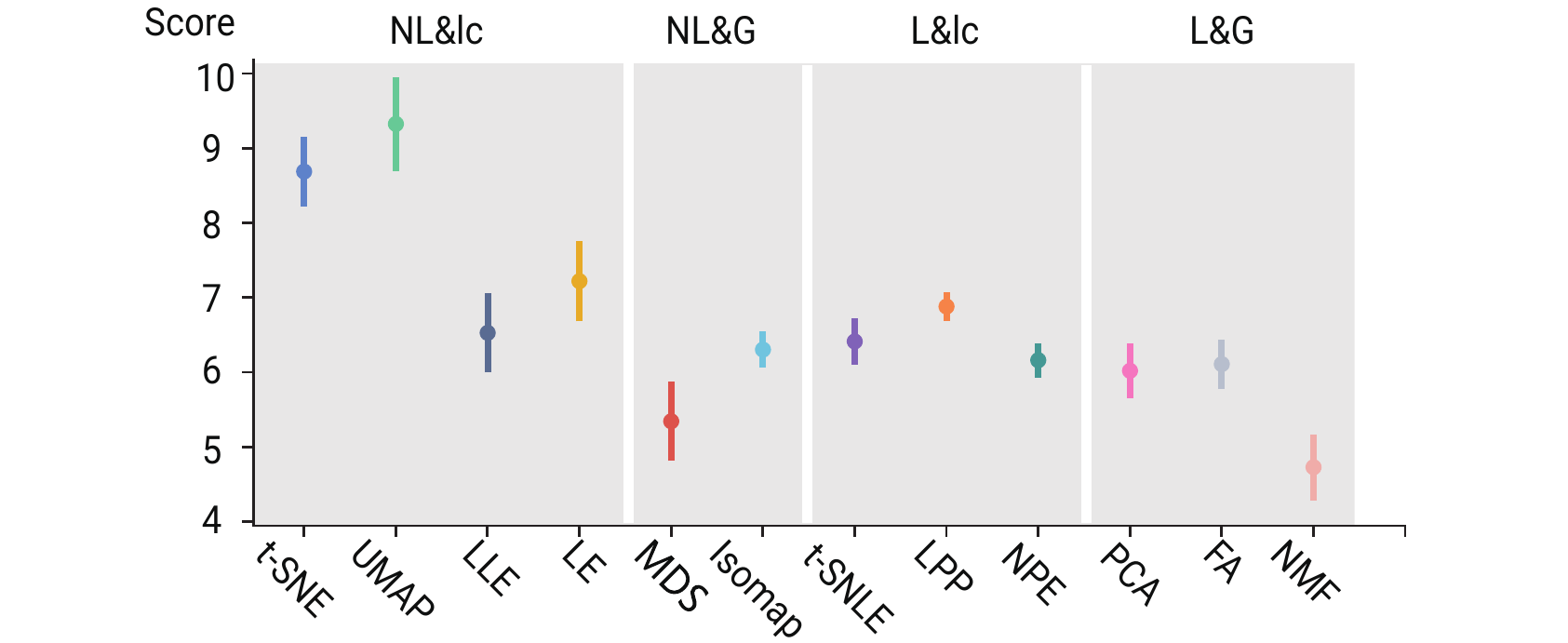}
\setlength{\abovecaptionskip}{0.cm}
\caption{Average ranking scores of \textbf{E5} (the higher the better). Error bars indicate 95\% confidence intervals.}
\label{fig:task5} 
\end{figure}

The results of \textbf{E5} are shown in Fig.~\ref{fig:task5}. The rankings of the most preferred techniques are: UMAP (9.26), t-SNE (8.60), LE (7.07), LPP (6.74), LLE (6.38), t-SNLE (6.26), Isomap (6.15), NPE (6.00),  FA (5.95), PCA (5.86), MDS (5.16), NMF (4.53). Local techniques obtain the top four preferred techniques with UMAP and t-SNE ranked the highest positions. It suggests a strong preference for the local techniques from the participants. Regarding linearity, LPP is ranked as the 4th preferred techniques. It indicates that there is no strong tendency toward a particular linearity type by the participants. 
It is worth noting that there could be misunderstandings in interpreting the projection results. For example, if a single cluster in the data space was scattered into two clusters by the projection, the participants could view it as a correct visual cluster. Similar to Lewis et al.\cite{lewis2012behavioral}, we made a correlation analysis between the results of the subjective experiment and the results of objective experiments. The correlation coefficients between the rankings of E5 and E1 are 0.73 (precision) and 0.76 (recall). The correlation confirms the consistency and differences between the subjective experiment and objective experiments.


\subsection{Summary}

Our experiment results indicate that there is no universally better DR technique and high-level property. Therefore, it is essential to choose appropriate techniques for enhancing particular tasks involving visual cluster analysis. 
(1) In cluster identification, local DR techniques perform better than global DR techniques in cluster identification if they have the same linearity type, except LLE;
non-linear DR techniques perform better than linear DR techniques if they have the same locality type, except LLE; t-SNE and UMAP have an outstanding performance.
(2) In membership identification, local DR techniques perform better than global DR techniques if they have the same linearity type, except LLE; non-linear DR techniques perform better than linear DR techniques if they have the same locality type, except LLE; t-SNE and UMAP have an outstanding performance.
(3) In density comparison, linear DR techniques perform better than non-linear DR techniques if they have the same locality type, except MDS; t-SNLE and NPE are the preffered techniques.
(4) In distance comparison, NMF and UMAP are the preffered techniques.
The experiment results shed light on DR techniques that received less attentions, such as NPE, t-SNLE.




\section{Discussion}

\begin{figure}[!tb]
\centering
\includegraphics[width=1\linewidth]{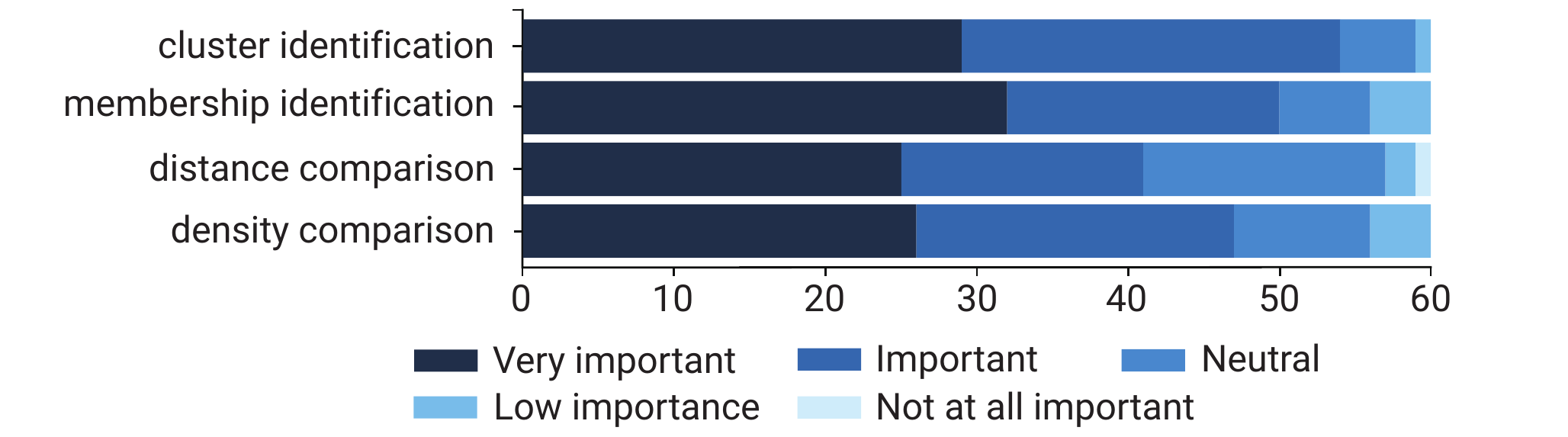}
\setlength{\abovecaptionskip}{0.cm}
\caption{The importance rating of the four visual cluster
analysis tasks evaluated in our experiments.}
\label{fig:sub2} 
\end{figure}

\noindent\textbf{Visual cluster analysis tasks.}
The participants answered the subjective questionnaire of the importance of the four analysis tasks after completing the experiments. 
As Fig.~\ref{fig:sub2} shows, the average scores of the four tasks are cluster identification (4.36), membership identification (4.30), density comparison (4.15), and cluster distance comparison (4.03), respectively.
These scores once again confirm that these four tasks are common concerns when using DR techniques for visual cluster analysis.
The experiment results also show that overall DR techniques perform better in E1 and E2 than in E3 and E4 in terms of precision and recall. 
The average precision of all DR techniques in E1--E4 are 0.69, 0.67, 0.32, and 0.29, respectively.The average recall of them are 0.67, 0.74, 0.60, and 0.46, respectively. It suggests a research opportunity to design DR techniques that support E3 and E4 better.

\noindent\textbf{The homogeneity of datasets.} We check the homogeneity of eight datasets by comparing the normalized standard deviations of precision, recall, and time for each task and each DR technique. As Table~\ref{tab:standard} shows, the normalized standard deviations are ranged from 0.03 to 0.46 and present a reasonable homogeneity of datasets. Another observation is that t-SNE and UMAP that perform better than other DR techniques in most tasks, also present low normalized standard deviations. Therefore, t-SNE and UMAP are recommended in the aspects of performance and stability on datasets with different characteristics.


\begin{table}[tb]     
\centering
\tiny
\setlength\tabcolsep{4pt}
\begin{tabular}{lrrrrrrrrrrrr} 
\toprule  
\ &E1P&E1R&E1T&E2P&E2R&E2T&E3P&E3R&E3T&E4P&E4R&E4T\\
\midrule            
t-SNE &  0.05 &  0.12 &  0.33 &  0.14 &  0.18 &  0.15 &  0.39 &  0.36 &  0.36 &  0.46 &  0.31 &  0.36 \\
UMAP &  0.03 &  0.13 &  0.15 &  0.14 &  0.17 &  0.1 &  0.35 &  0.32 &  0.34 &  0.12 &  0.05 &  0.39 \\
LE &  0.26 &  0.27 &  0.26 &  0.23 &  0.15 &  0.28 &  0.41 &  0.38 &  0.19 &  0.36 &  0.31 &  0.29 \\
LLE &  0.31 &  0.29 &  0.38 &  0.41 &  0.35 &  0.28 &  0.34 &  0.38 &  0.16 &  0.25 &  0.37 &  0.32 \\
MDS &  0.28 &  0.26 &  0.16 &  0.31 &  0.34 &  0.27 &  0.27 &  0.31 &  0.24 &  0.37 &  0.35 &  0.35 \\
Isomap &  0.26 &  0.26 &  0.28 &  0.32 &  0.37 &  0.34 &  0.35 &  0.36 &  0.31 &  0.42 &  0.39 &  0.24 \\
t-SNLE &  0.35 &  0.34 &  0.32 &  0.42 &  0.34 &  0.27 &  0.35 &  0.39 &  0.39 &  0.35 &  0.36 &  0.37 \\
LPP &  0.15 &  0.12 &  0.26 &  0.24 &  0.35 &  0.22 &  0.27 &  0.36 &  0.19 &  0.35 &  0.44 &  0.36 \\
NPE &  0.32 &  0.33 &  0.15 &  0.35 &  0.44 &  0.27 &  0.38 &  0.35 &  0.24 &  0.3 &  0.36 &  0.27 \\
PCA &  0.23 &  0.25 &  0.34 &  0.32 &  0.37 &  0.26 &  0.33 &  0.35 &  0.37 &  0.34 &  0.32 &  0.35 \\
FA &  0.3 &  0.3 &  0.34 &  0.42 &  0.43 &  0.26 &  0.3 &  0.35 &  0.32 &  0.36 &  0.36 &  0.2 \\
NMF &  0.27 &  0.25 &  0.19 &  0.26 &  0.32 &  0.25 &  0.37 &  0.26 &  0.23 &  0.34 &  0.27 &  0.31 \\
\bottomrule         
\end{tabular}
\renewcommand\tablename{Table}
\caption{The homogeneity of eight datasets. E\#P, E\#R, and E\#T represent the normalized standard deviations of precision, recall, and completion time in E\# in terms of eight datasets, respectively.}
\label{tab:standard}
\end{table}



\noindent\textbf{Cluster identification and class separation. }In this study, our evaluation is based on the perception of cluster identification. Participants were provided with monochrome scatterplots without color-coded class information. It is worth noting that we have also used the class label information with the datasets as the ground truth of clusters.
Therefore, in selecting the datasets, we check the visual separation of clusters with the visual class separation measure. However, when a class is scattered into several clusters, the class separation measure also presents a good but misleading result. This issue can be addressed by additional user controlled selection with colored scatterplot. More discussion on this issue is referred to Aupetit~\cite{aupetit2014sanity}.

\noindent\textbf{Parameter robustness of DR techniques.}
Our pre-study provides an opportunity to analyze the DR techniques in the aspect of parameter robustness.
In our pre-study, participants were asked to select the result that best fits the structure of datasets among five settings.
If the selections are uniformly distributed in all five settings, we can infer that the tuning of parameters has few effects on the quality of projection.
Otherwise, a big variation of the setting distribution indicates that the DR technique is sensitive to the parameter.
We calculated the average standard deviation across datasets of each tested eight DR techniques in the pre-study.
The rankings are: t-SNLE (3.25), t-SNE (3.15), LLE (3.11), UMAP (3.04), LE (2.57), NPE (2.39), Isomap (2.35), and LPP (1.85).
The results show that although t-SNE is one of the most popular techniques, its sensitivity to the \textit{perplexity} parameter should be considered. This observation is the same as Espadoto et al.~\cite{espadoto2019towards}.
It is also worth noting that UMAP is less sensitive than t-SNE while having similar performance in most tasks.

\noindent\textbf{Conducting remote studies. }
Remote study lacks a face-to-face communication environment and yields several challenges for a control study. First, the test should be performed independently. In our study, participants are required to set an independent environment to avoid possible disturbance.
Second, the test could be interrupted due to network issues. We ask participants to test their network to ensure a stable connection. During the test, if a participant is disconnected for more than ten minutes, we discard this sample to ensure the validity of tests. Third, the remote communication would be limited in a remote study. To address this issue, we monitor the experimental screen of participants through screen sharing throughout the test. We also communicate with participant during the test in voice.

\noindent\textbf{Order of experiments and trials.}
A limitation of our study is the order of experiments and trials. Because E3 is conducted before E4, it may reveal a part of ground truth clusters and make a potential threat to validity of E4. If the given cluster in E3 happens to be the cluster with the highest density in E4: 1) when several classes overlap in a single cluster, then an increase of precision would occur; 2) when a single class is scattered in several clusters, then an increase of recall would occur. In addition, learning effects should be considered. A Latin square design could address this issue.

\noindent\textbf{Limitations and future work.}
Due to the different characteristics of datasets, DR techniques will inevitably produce different performance in different types of datasets. Therefore, the experimental results may have a certain correlation with the datasets. In this study, we focus on the characteristics of locality and linearity and four objective visual clustering tasks. In the future, it is valuable to perform further studies on other characteristics (for example, false and missed neighbors could have an impact on clustering tasks that is worth evaluating~\cite{nonato2018multidimensional} ) and tasks. The linear-nonlinear pairs of techniques, such as t-SNE and t-SNLE, also shed light on the development of DR techniques and design of comparative analysis.

\section{Conclusion}

In this paper, we present an empirical evaluation of DR techniques from the perspective of perception in visual cluster analysis tasks. 
Twelve representative DR techniques are identified from a literature review. They are grouped into groups based on linearity and locality. 
We first conduct a pre-study to determine the proper input parameters for each DR technique. Then, the techniques are formally evaluated in four controlled experiments each of which focuses on hypotheses on a particular analytical task. A subjective task is also conducted to collect users' preferences on each technique. 
Several guidelines and interesting insights are provided by the result analysis.


\acknowledgments{We would like to thank the helpful comments from the anonymous reviewers. This work is supported by the National Natural Science Foundation of China (No. 61872389).}

\bibliographystyle{abbrv-doi}

\bibliography{template}

\begin{thebibliography}{10}

\bibitem{abbas2019clustme}
M.~M. Abbas, M.~Aupetit, M.~Sedlmair, and H.~Bensmail.
\newblock Clustme: A visual quality measure for ranking monochrome scatterplots
  based on cluster patterns.
\newblock {\em Computer Graphics Forum}, 38(3):225--236, 2019.

\bibitem{arora2018analysis}
S.~Arora, W.~Hu, and P.~K. Kothari.
\newblock An analysis of the t-sne algorithm for data visualization.
\newblock In {\em Proceedings of Conference on Learning Theory}, pp.
  1455--1462, 2018.

\bibitem{aupetit2014sanity}
M.~Aupetit.
\newblock Sanity check for class-coloring-based evaluation of dimension
  reduction techniques.
\newblock In {\em Proceedings of the Fifth Workshop on Beyond Time and Errors:
  Novel Evaluation Methods for Visualization}, pp. 134--141, 2014.

\bibitem{aupetit2002gamma}
M.~Aupetit, P.~Couturier, and P.~Massotte.
\newblock $\gamma$-observable neighbours for vector quantization.
\newblock {\em Neural networks}, 15(8-9):1017--1027, 2002.

\bibitem{aupetit2016sepme}
M.~Aupetit and M.~Sedlmair.
\newblock Sepme: 2002 new visual separation measures.
\newblock In {\em Proceedings of IEEE Pacific Visualization Symposium}, pp.
  1--8, 2016.

\bibitem{aupetit2019toward}
M.~Aupetit, M.~Sedlmair, M.~M. Abbas, A.~Baggag, and H.~Bensmail.
\newblock Toward perception-based evaluation of clustering techniques for
  visual analytics.
\newblock In {\em Proceedings of IEEE Visualization Conference}, pp. 141--145,
  2019.

\bibitem{becht2019dimensionality}
E.~Becht, L.~McInnes, J.~Healy, C.-A. Dutertre, I.~W. Kwok, L.~G. Ng,
  F.~Ginhoux, and E.~W. Newell.
\newblock Dimensionality reduction for visualizing single-cell data using umap.
\newblock {\em Nature biotechnology}, 37(1):38--44, 2019.

\bibitem{belkin2002laplacian}
M.~Belkin and P.~Niyogi.
\newblock Laplacian eigenmaps and spectral techniques for embedding and
  clustering.
\newblock In {\em Proceedings of Conference on Neural Information Processing
  Systems}, pp. 585--591, 2002.

\bibitem{bunte2012general}
K.~Bunte, M.~Biehl, and B.~Hammer.
\newblock A general framework for dimensionality-reducing data visualization
  mapping.
\newblock {\em Neural Computation}, 24(3):771--804, 2012.

\bibitem{cai2008non}
D.~Cai, X.~He, X.~Wu, and J.~Han.
\newblock Non-negative matrix factorization on manifold.
\newblock In {\em Proceedings of IEEE International Conference on Data Mining},
  pp. 63--72, 2008.

\bibitem{chatzimparmpas2020t}
A.~Chatzimparmpas, R.~M. Martins, and A.~Kerren.
\newblock t-visne: Interactive assessment and interpretation of t-sne
  projections.
\newblock {\em IEEE transactions on visualization and computer graphics},
  26(8):2696--2714, 2020.

\bibitem{cheng2015data}
S.~Cheng and K.~Mueller.
\newblock The data context map: Fusing data and attributes into a unified
  display.
\newblock {\em IEEE transactions on visualization and computer graphics},
  22(1):121--130, 2015.

\bibitem{de2003global}
V.~De~Silva and J.~B. Tenenbaum.
\newblock Global versus local methods in nonlinear dimensionality reduction.
\newblock In {\em Proceedings of Neural Information Processing Systems}, pp.
  721--728, 2003.

\bibitem{ding2004k}
C.~Ding and X.~He.
\newblock K-means clustering via principal component analysis.
\newblock In {\em Proceedings of International Conference on Machine Learning},
  p.~29, 2004.

\bibitem{Dua2019uci}
D.~Dua and C.~Graff.
\newblock {UCI} machine learning repository, 2017.

\bibitem{espadoto2019towards}
M.~Espadoto, R.~M. Martins, A.~Kerren, N.~S. Hirata, and A.~C. Telea.
\newblock Towards a quantitative survey of dimension reduction techniques.
\newblock {\em IEEE Transactions on Visualization and Computer Graphics},
  27(3):2153--2173, 2021.

\bibitem{etemadpour2014perception}
R.~Etemadpour, R.~Motta, J.~G. de~Souza~Paiva, R.~Minghim, M.~C.~F.
  de~Oliveira, and L.~Linsen.
\newblock Perception-based evaluation of projection methods for
  multidimensional data visualization.
\newblock {\em IEEE Transactions on Visualization and Computer Graphics},
  21(1):81--94, 2014.

\bibitem{forina1983classification}
M.~Forina, C.~Armanino, S.~Lanteri, and E.~Tiscornia.
\newblock Classification of olive oils from their fatty acid composition.
\newblock In {\em Proceedings of Food research and data analysis}, pp.
  189--214, 1983.

\bibitem{garcia2013stability}
F.~J. Garc{\'\i}a~Fern{\'a}ndez, M.~Verleysen, J.~A. Lee, and
  I.~D{\'\i}az~Blanco.
\newblock Stability comparison of dimensionality reduction techniques attending
  to data and parameter variations.
\newblock In {\em Proceedings of Eurographics Conference on Visualization},
  2013.

\bibitem{georghiades2001few}
A.~S. Georghiades, P.~N. Belhumeur, and D.~J. Kriegman.
\newblock From few to many: Illumination cone models for face recognition under
  variable lighting and pose.
\newblock {\em IEEE transactions on pattern analysis and machine intelligence},
  23(6):643--660, 2001.

\bibitem{han2019visual}
Q.~Han, D.~Thom, M.~John, S.~Koch, F.~Heimerl, and T.~Ertl.
\newblock Visual quality guidance for document exploration with focus+ context
  techniques.
\newblock {\em IEEE transactions on visualization and computer graphics},
  26(8):2715--2731, 2019.

\bibitem{he2005neighborhood}
X.~He, D.~Cai, S.~Yan, and H.-J. Zhang.
\newblock Neighborhood preserving embedding.
\newblock In {\em Proceedings of International Conference on Computer Vision},
  pp. 1208--1213, 2005.

\bibitem{he2004locality}
X.~He and P.~Niyogi.
\newblock Locality preserving projections.
\newblock In {\em Proceedings of Neural Information Processing Systems}, pp.
  153--160, 2004.

\bibitem{ingram2008glimmer}
S.~Ingram, T.~Munzner, and M.~Olano.
\newblock Glimmer: Multilevel mds on the gpu.
\newblock {\em IEEE Transactions on Visualization and Computer Graphics},
  15(2):249--261, 2008.

\bibitem{joia2011local}
P.~Joia, D.~Coimbra, J.~A. Cuminato, F.~V. Paulovich, and L.~G. Nonato.
\newblock Local affine multidimensional projection.
\newblock {\em IEEE Transactions on Visualization and Computer Graphics},
  17(12):2563--2571, 2011.

\bibitem{jolliffe1986principal}
I.~T. Jolliffe.
\newblock Principal components in regression analysis.
\newblock In {\em Principal component analysis}, pp. 129--155. Springer, 1986.

\bibitem{koren2003visualization}
Y.~Koren and L.~Carmel.
\newblock Visualization of labeled data using linear transformations.
\newblock In {\em Proceedings of IEEE Symposium on Information Visualization},
  pp. 121--128, 2003.

\bibitem{kruskal1978multidimensional}
J.~B. Kruskal and M.~Wish.
\newblock {\em Multidimensional scaling}, vol.~11.
\newblock Sage, 1978.

\bibitem{lee1999learning}
D.~D. Lee and H.~S. Seung.
\newblock Learning the parts of objects by non-negative matrix factorization.
\newblock {\em Nature}, 401(6755):788--791, 1999.

\bibitem{lewis2012human}
J.~Lewis, M.~Ackerman, and V.~de~Sa.
\newblock Human cluster evaluation and formal quality measures: A comparative
  study.
\newblock In {\em Proceedings of the Annual Meeting of the Cognitive Science
  Society}, pp. 1870--1875, 2012.

\bibitem{lewis2012behavioral}
J.~Lewis, L.~Van~der Maaten, and V.~de~Sa.
\newblock A behavioral investigation of dimensionality reduction.
\newblock In {\em Proceedings of the Annual Meeting of the Cognitive Science
  Society}, pp. 671--–676, 2012.

\bibitem{linderman2019clustering}
G.~C. Linderman and S.~Steinerberger.
\newblock Clustering with t-sne, provably.
\newblock {\em SIAM Journal on Mathematics of Data Science}, 1(2):313--332,
  2019.

\bibitem{liu2017towards}
M.~Liu, J.~Shi, Z.~Li, C.~Li, J.~Zhu, and S.~Liu.
\newblock Towards better analysis of deep convolutional neural networks.
\newblock {\em IEEE Transactions on Visualization and Computer Graphics},
  23(1):91--100, 2017.

\bibitem{mair2018factor}
P.~Mair.
\newblock Factor analysis.
\newblock In {\em Modern Psychometrics with R}, pp. 17--61. Springer, 2018.

\bibitem{martins2014visual}
R.~M. Martins, D.~B. Coimbra, R.~Minghim, and A.~C. Telea.
\newblock Visual analysis of dimensionality reduction quality for parameterized
  projections.
\newblock {\em Computers \& Graphics}, 41:26--42, 2014.

\bibitem{mcinnes2018umap}
L.~McInnes, J.~Healy, and J.~Melville.
\newblock Umap: Uniform manifold approximation and projection for dimension
  reduction.
\newblock {\em arXiv preprint arXiv:1802.03426}, 2018.

\bibitem{narayan2021assessing}
A.~Narayan, B.~Berger, and H.~Cho.
\newblock Assessing single-cell transcriptomic variability through
  density-preserving data visualization.
\newblock {\em Nature Biotechnology}, pp. 1--10, 2021.

\bibitem{nonato2018multidimensional}
L.~G. Nonato and M.~Aupetit.
\newblock Multidimensional projection for visual analytics: Linking techniques
  with distortions, tasks, and layout enrichment.
\newblock {\em IEEE Transactions on Visualization and Computer Graphics},
  25(8):2650--2673, 2018.

\bibitem{paulovich2008least}
F.~V. Paulovich, L.~G. Nonato, R.~Minghim, and H.~Levkowitz.
\newblock Least square projection: A fast high-precision multidimensional
  projection technique and its application to document mapping.
\newblock {\em IEEE Transactions on Visualization and Computer Graphics},
  14(3):564--575, 2008.

\bibitem{prajapati2010sample}
B.~Prajapati, M.~Dunne, and R.~Armstrong.
\newblock Sample size estimation and statistical power analyses.
\newblock {\em Optometry today}, 16(7):10--18, 2010.

\bibitem{rousseeuw1987silhouettes}
P.~J. Rousseeuw.
\newblock Silhouettes: a graphical aid to the interpretation and validation of
  cluster analysis.
\newblock {\em Journal of computational and applied mathematics}, 20:53--65,
  1987.

\bibitem{roweis2000nonlinear}
S.~T. Roweis and L.~K. Saul.
\newblock Nonlinear dimensionality reduction by locally linear embedding.
\newblock {\em science}, 290(5500):2323--2326, 2000.

\bibitem{schreck2010techniques}
T.~Schreck, T.~Von~Landesberger, and S.~Bremm.
\newblock Techniques for precision-based visual analysis of projected data.
\newblock {\em Information Visualization}, 9(3):181--193, 2010.

\bibitem{sedlmair2012taxonomy}
M.~Sedlmair, A.~Tatu, T.~Munzner, and M.~Tory.
\newblock A taxonomy of visual cluster separation factors.
\newblock {\em Computer Graphics Forum}, 31(3pt4):1335--1344, 2012.

\bibitem{spathis2018fast}
D.~Spathis, N.~Passalis, and A.~Tefas.
\newblock Fast, visual and interactive semi-supervised dimensionality
  reduction.
\newblock In {\em Proceedings of the European Conference on Computer Vision},
  pp. 550--563, 2018.

\bibitem{tenenbaum2000global}
J.~B. Tenenbaum, V.~De~Silva, and J.~C. Langford.
\newblock A global geometric framework for nonlinear dimensionality reduction.
\newblock {\em science}, 290(5500):2319--2323, 2000.

\bibitem{maaten2008visualizing}
L.~Van~der Maaten and G.~Hinton.
\newblock Visualizing data using t-sne.
\newblock {\em Journal of machine learning research}, 9(86):2579--2605, 2008.

\bibitem{van2009dimensionality}
L.~Van Der~Maaten, E.~Postma, and J.~Van~den Herik.
\newblock Dimensionality reduction: a comparative review.
\newblock {\em J Mach Learn Res}, 10(66-71):13, 2009.

\bibitem{ventocilla2020comparative}
E.~Ventocilla and M.~Riveiro.
\newblock A comparative user study of visualization techniques for cluster
  analysis of multidimensional data sets.
\newblock {\em Information Visualization}, 19(4):318--338, 2020.

\bibitem{vernier2020quantitative}
E.~F. Vernier, R.~Garcia, I.~d. Silva, J.~L.~D. Comba, and A.~C. Telea.
\newblock Quantitative evaluation of time-dependent multidimensional projection
  techniques.
\newblock In {\em Computer Graphics Forum}, pp. 241--252, 2020.

\bibitem{wang2017perception}
Y.~Wang, K.~Feng, X.~Chu, J.~Zhang, C.-W. Fu, M.~Sedlmair, X.~Yu, and B.~Chen.
\newblock A perception-driven approach to supervised dimensionality reduction
  for visualization.
\newblock {\em IEEE transactions on visualization and computer graphics},
  24(5):1828--1840, 2017.

\bibitem{wattenberg2016use}
M.~Wattenberg, F.~Vi{\'e}gas, and I.~Johnson.
\newblock How to use t-sne effectively.
\newblock {\em Distill}, 1(10):e2, 2016.

\bibitem{wenskovitch2017towards}
J.~Wenskovitch, I.~Crandell, N.~Ramakrishnan, L.~House, and C.~North.
\newblock Towards a systematic combination of dimension reduction and
  clustering in visual analytics.
\newblock {\em IEEE transactions on visualization and computer graphics},
  24(1):131--141, 2017.

\bibitem{wilkinson2008scagnostics}
L.~Wilkinson and G.~Wills.
\newblock Scagnostics distributions.
\newblock {\em Journal of Computational and Graphical Statistics},
  17(2):473--491, 2008.

\bibitem{wold1987principal}
S.~Wold, K.~Esbensen, and P.~Geladi.
\newblock Principal component analysis.
\newblock {\em Chemometrics and intelligent laboratory systems}, 2(1-3):37--52,
  1987.

\bibitem{xiang2019interactive}
S.~Xiang, X.~Ye, J.~Xia, J.~Wu, Y.~Chen, and S.~Liu.
\newblock Interactive correction of mislabeled training data.
\newblock In {\em Proceedings of IEEE Conference on Visual Analytics Science
  and Technology}, pp. 57--68, 2019.

\bibitem{xu2017evaluating}
K.~Xu, L.~Zhang, D.~P{\'e}rez, P.~H. Nguyen, and A.~Ogilvie-Smith.
\newblock Evaluating interactive visualization of multidimensional data
  projection with feature transformation.
\newblock {\em Multimodal Technologies and Interaction}, 1(3):13, 2017.

\bibitem{yuan2021survey}
J.~Yuan, C.~Chen, W.~Yang, M.~Liu, J.~Xia, and S.~Liu.
\newblock A survey of visual analytics techniques for machine learning.
\newblock {\em Computational Visual Media}, 7(1):3--36, 2021.

\bibitem{yuan2020evaluation}
J.~Yuan, S.~Xiang, J.~Xia, L.~Yu, and S.~Liu.
\newblock Evaluation of sampling methods for scatterplots.
\newblock {\em IEEE Transactions on Visualization and Computer Graphics},
  27(2):1720--1730, 2020.

\bibitem{yue2019sportfolio}
X.~Yue, J.~Bai, Q.~Liu, Y.~Tang, A.~Puri, K.~Li, and H.~Qu.
\newblock s{P}ortfolio: Stratified visual analysis of stock portfolios.
\newblock {\em IEEE Transactions on Visualization and Computer Graphics},
  26(1):601--610, 2019.

\bibitem{zhao2018evaluating}
Y.~Zhao, F.~Luo, M.~Chen, Y.~Wang, J.~Xia, F.~Zhou, Y.~Wang, Y.~Chen, and
  W.~Chen.
\newblock Evaluating multi-dimensional visualizations for understanding fuzzy
  clusters.
\newblock {\em IEEE transactions on visualization and computer graphics},
  25(1):12--21, 2018.

\end{thebibliography}
\end{document}